\documentclass[twocolumn]{aastex631}
\usepackage{CJK}
\usepackage{fontawesome}
\usepackage{amsmath}
\usepackage{booktabs}
\usepackage{multirow}
\accepted{March 12, 2024}

\submitjournal{ApJ}

\begin{document}

\title{Spectroscopic Observations of the Solar Corona during the 2017 August 21 Total Solar Eclipse: Comparison of Spectral Line Widths and Doppler Shifts
\\Between Open and Closed Magnetic Structures} 

\begin{CJK*}{UTF8}{gbsn}
\author[0000-0003-3908-1330]{Yingjie Zhu (朱英杰)}
\altaffiliation{DKIST Ambassador.  \\
Now at Physikalisch Meteorologische Observatorium Davos, World Radiation Center (PMOD/WRC), 7260 Davos, Switzerland}
\affiliation{Department of Climate and Space Sciences and Engineering, University of Michigan, \\
Ann Arbor, MI 48109, USA}

\author[0000-0003-4089-9316]{Shadia R. Habbal}
\affiliation{Institute for Astronomy, University of Hawaii, 2680 Woodlawn Drive, Honolulu, HI, USA}

\author[0000-0001-6573-7810]{Adalbert Ding}
\affiliation{Institute of Optics and Atomic Physics, Technische Universit\"at Berlin, Berlin, Germany}
\affiliation{Institut f\"ur Technische Physik, Berlin, Germany}

\author{Bryan Yamashiro}
\affiliation{Institute for Astronomy, University of Hawaii, 2680 Woodlawn Drive, Honolulu, HI, USA}

\author[0000-0002-9325-9884]{Enrico Landi}
\affiliation{Department of Climate and Space Sciences and Engineering, University of Michigan, \\
Ann Arbor, MI 48109, USA}

\author[0000-0002-6396-8209]{Benjamin Boe}
\affiliation{Wentworth Institute of Technology, Boston, MA 02115, USA}
\affiliation{Institute for Astronomy, University of Hawaii, 2680 Woodlawn Drive, Honolulu, HI, USA}

\author[0000-0002-8937-5620]{Sage Constantinou}
\affiliation{Institute for Astronomy, University of Hawaii, 2680 Woodlawn Drive, Honolulu, HI, USA}

\author[0009-0007-8518-1726]{Michael Nassir}
\affiliation{Institute for Astronomy, University of Hawaii, 2680 Woodlawn Drive, Honolulu, HI, USA}

\correspondingauthor{Yingjie Zhu (朱英杰)}
\email{yingjie.zhu@pmodwrc.ch}



\begin{abstract}

The spectroscopic observations presented here were acquired during the 2017 August 21 total solar eclipse with a three-channel partially multiplexed imaging spectrometer (3PAMIS) operating at extremely high orders ($>$ 50). The 4\,$R_\odot$ extent of the slit in the North-South direction scanned the corona starting from the central meridian out to approximately 1.0\,$R_\odot$ off the east limb throughout totality. The line widths and Doppler shifts of the Fe \textsc{x} (637.4\,nm) and Fe \textsc{xiv} (530.3\,nm) emission lines, characteristic of $1.1 \times 10^6$ K and  $1.8 \times 10^6$ K electron temperatures respectively, varied across the different coronal structures intercepted by the slit. Fe \textsc{xiv} was the dominant emission in the closed fields of an active region and the base of a streamer, with relatively constant 20 - 30\,km\,s$^{-1}$ line widths independent of the height. In contrast, Fe \textsc{x} emission exhibited broader ($>40$\,km\,s$^{-1}$) line widths in open fields which increased with height, in particular in the polar coronal hole. Inferences of line widths and Doppler shifts were consistent with extreme ultraviolet (EUV) observations from Hinode/EIS, as well as with the near-infrared Fe \textsc{xiii} 1074\,nm line observed by CoMP. The differences in the spectral line widths between distinct coronal structures are interpreted as an indication of the predominance of wave heating in open structures versus localized heating in closed structures. This study underscores the unparalleled advantages and the enormous potential of TSE spectroscopy in measuring line widths simultaneously in open and closed fields at high altitudes, with minimal exposure times, stray light levels, and instrumental widths.

\end{abstract}

\keywords{Total eclipses (1704) --- Spectroscopy (1558) --- Solar coronal lines (2038) --- Solar coronal streamers (1486) --- Quiet solar corona (1992) --- Solar coronal holes (1484)}


\section{Introduction} \label{sec:intro}

Total solar eclipse (TSE) spectroscopic observations of the `green' line in 1869 by Young and Harkness led to the discovery of a $1.8 \times 10^6$ K coronal electron temperature,  when its correct identification as Fe \textsc{xiv} emission at 530.3\,nm was made by \cite{Grotrian1939} and \cite{ Edlen1943}. Following this seminal discovery, TSE spectroscopic observations have been pursued in earnest. They led to the discovery of a rich coronal spectrum with different ionization states of elements, such as Ni, Ar, and Ca, to name a few \citep[see][]{Jefferies1971}. In addition to inferences of the electron temperature \citep[e.g.,][]{Boe2022,Boe2023}, and chemical composition, spectral lines offer fundamental diagnostic tools such as inferences of the ion effective temperature \citep{DelZanna2018}, which include contributions from ion temperatures and nonthermal motions along the line of sight. Doppler shifts, when present, yield mass motions, both steady and dynamic.

The list of identified emission lines in the early TSE spectroscopic observations did not always report the same emission lines. Furthermore, decades of spectroscopic observations also differed in the observed line widths and their variations across the corona,\citep{Kim2000,Koutchmy2005,Raju2011,Prabhakar2019}. 
 These differences can be readily accounted for by differences in the underlying structures, covering a range of electron temperatures, as resolved by complementary imaging observations of coronal emission lines during totality \citep[e.g.,][]{Habbal2011,Habbal2021,Boe2018}.

Despite their paucity, TSE spectral and imaging data remain unique scientific assets for exploring the properties of the different manifestations of coronal heating and solar wind acceleration mechanisms responsible for these observables. The uniqueness of these observations stems from the properties of the emission from coronal forbidden lines, which is dominated by radiative excitation \citep{Habbal2007}. This property enables the detection of the emission out to much larger distances than extreme ultraviolet imaging and spectroscopy, as the latter lines are dominated by collisional excitation, hence detectable only at shorter distances.

This paper presents an analysis of spectroscopic observations of the Fe \textsc{x} 637.4\,nm and Fe \textsc{xiv} 530.3\,nm lines obtained during the 2017 August 21 TSE. They capitalize on the distinct advantage of the spatial extent of emission from coronal forbidden lines spanning at least 1\,$R_\odot$ above the limb, thus exploring a range of different coronal structures. The primary focus of the analysis is on the spectral line widths from which the effective ion temperature can be inferred, and on any Doppler shifts when present. The spectral observations are placed in the context of emission line imaging of Fe \textsc{xi} 789.2\,nm and Fe \textsc{xiv} 530.3\,nm acquired at the same time \citep{Boe2020}. The eclipse observations are complemented by
spectroscopic observations of extreme ultraviolet (EUV) lines taken with the EUV Imaging Spectrograph \citep[EIS;][]{Culhane2007} on board the Hinode spacecraft \citep{Kosugi2007} as well as in the Fe \textsc{xiii} 1074.7\,nm near-infrared line with the ground-based Coronal Multichannel Polarimeter \citep[CoMP;][]{Tomczyk2008}.
 The observations, including methodology and the specifics of the spectrometer, are given in Section~\ref{sec:PaMIS_results}. The ancillary space-based and ground-based observations are presented in Section~\ref{sec:space_ground}. This is followed by a discussion including comparisons between the different instruments in Section~\ref{sec:dis}.  Concluding remarks with a summary of the outstanding findings and their implications are given in Section~\ref{sec:sum}.

\end{CJK*}

\section{3PAMIS Observations and Results}\label{sec:PaMIS_results}
\subsection{Operation and Data Acquisition}

During the 2017 TSE, spectroscopic observations were obtained using a three-channel PArtially Multiplexed Imaging Spectrometer (3PAMIS).  The 3PAMIS has a design similar to the dual-channel (2PAMIS) spectrometer used at the 2015 TSE \citep{Ding2017}.

With this spectrometer, the eclipsed Sun is imaged onto a slit mirror using a tele lens (NIKON ED Nikkor $f=300$\,mm, F/2.8). The transmitted light is made parallel by a collimator lens (ASKANIA Askinar $f=100$\,mm, F/1.9), and then passed through three dichroic mirrors which separate the spectrum into three wavelength bands: blue(400--500\,nm), green(500--610\,nm), and red(610--1100\,nm). The light from each of these regions is dispersed by three diffraction gratings into different output angles, depending on the wavelength and the diffraction order. Each of the three beams is focused onto a CCD camera (ATIK Infinity) with a lens system (NIKON Nikkor $f=50$\,mm, F/1.8). Schott color filters (cut-off and bandpass filters) are used to correct the limitations of the dichroic mirrors.

A monitor camera captures the solar image reflected by the slit mirror and determines the slit position with respect to the Sun. Our analysis of the detector images showed that the chromium slit mirror coating is slightly transparent, resulting in ghost images of the solar limb and of bright prominences which are superimposed onto the coronal spectrum. 


\begin{figure}[htb!]
    \centering
    \includegraphics[width=\linewidth]{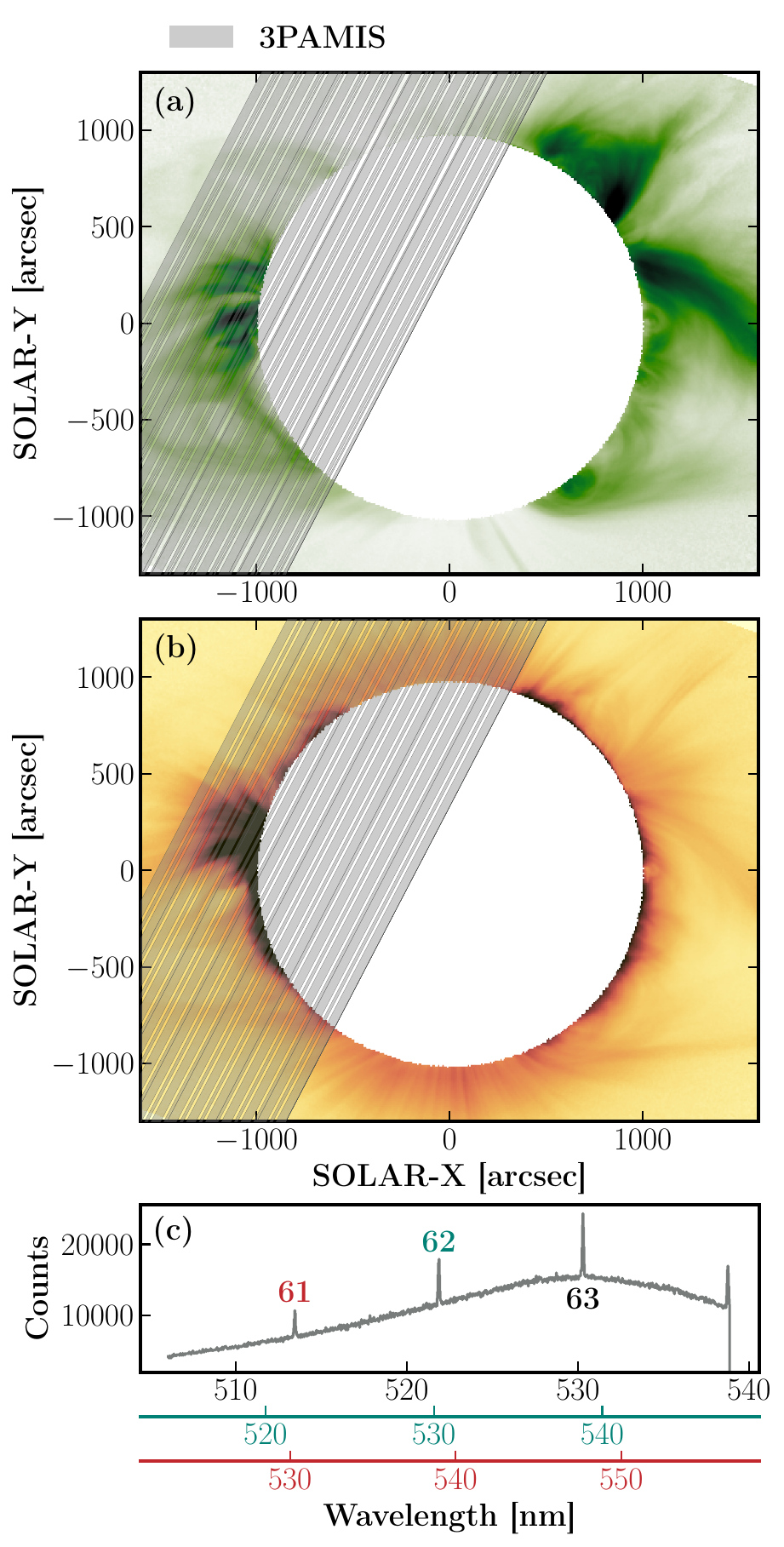}
    \caption{Overview of the 3PAMIS observation during the 2017 TSE and two Hinode/EIS observations made on 2017 August 21. (a) 3PAMIS/Green FOV (gray)
    overplotted on the Fe \textsc{xiv} 530.3\,nm line-to-continuum ratio image. The darker gray lines indicate the start and end times of each exposure. (b) 3PAMIS/Red FOV (gray)
    overplotted on the Fe \textsc{xi} 789.2\,nm line-to-continuum ratio image. (c) An example streamer spectrum showing Fe \textsc{xiv} 530.3\,nm line at the 61st, 62nd, and 63rd orders. Link to the \texttt{Jupyter} notebook creating this figure: \href{https://yjzhu-solar.github.io/Eclipse2017/ipynb_html/obs_summary.html}{\faGithub}.}
    \label{fig:obs_summray}
\end{figure}


The emission of Fe \textsc{xiv} (530.3\,nm) is observed in the green channel, and that of Fe \textsc{x} (637.4\,nm) is in the red channel along with other spectral lines (H, He, Fe \textsc{xi},...), all in very high orders. The Fe \textsc{xiv} emission is captured from 60\textsuperscript{th} to 64\textsuperscript{th} orders, while the Fe \textsc{x} 637.4\,nm line is recorded from 51\textsuperscript{st} to 53\textsuperscript{rd} orders. In addition to the spectral lines, the electron K-corona and F-corona continuum were also observed \citep{Boe2021}. However, the stacking of multiple orders of the white light continuum on the detector made its interpretation challenging. The wavelength scale of the green detector is approximately 0.025\,nm\,px$^{-1}$, and the red detector has a wavelength scale of around 0.030\,nm\,px$^{-1}$, yielding a resolution power $R\sim 20,000$.  

The 3PAMIS spectra covered a region corresponding to 4 $R_\odot$ along the slit direction. The pixel size is equivalent to 8\farcs3 in the spatial dimension. The spatial resolution perpendicular to the slit depends on the exposure time of each raster. This is because 3PAMIS made a sit-and-stare observation, with the Sun gradually moving across the slit as time went by.   

During the eclipse, the 3PAMIS data presented here were acquired at Guernsey State Park, Wyoming, USA, at 42\arcdeg18\farcm585, W 104\arcdeg47\farcm206, and an altitude of 1406\,m. Totality began at 17:45:37 UT (second contact), when the Sun was 55\fdg4 above the horizon, and ended at 17:47:56 UT (third contact). At the start of the totality, the slit was initially placed along the central meridian, tilted slightly from the solar northwest to southeast. The tracking motion of the mount was then disabled, and the Sun slowly drifted across the slit, enabling the acquisition of the coronal spectrum above the east limb. 3PAMIS made intermittent exposures when the slit scanned the east limb, leaving data gaps between each exposure due to the finite readout and download time of the two CCD detectors (see Figure~\ref{fig:obs_summray}a and b). The typical exposure times of the green detector were repeating sequentially at 0.5\,s, 1\,s, and 3\,s, while for the red detector, they were 1\,s and 3\,s.

The 3PAMIS data were corrected and calibrated through dark frame subtraction, curvature correction, and flat-fielding. In addition, the wavelength calibration was performed and the instrumental broadening was measured from the calibration frames taken in the laboratory. The 3PAMIS pointing was determined using the slit position in the context images and was coaligned with the narrow-bandpass images. The details of data reduction and calibration procedures are presented in Appendix~\ref{app:data_calib}. 

Figure~\ref{fig:obs_summray}c shows an example spectrum from the green detector, including the Fe \textsc{xiv} 530.3\,nm line at 61st, 62nd, and 63rd orders and the stacked multi-order continuum. The ambient continuum was first removed by a linear fit and followed by a single-Gaussian fit to the strongest orders of Fe \textsc{xiv} 530.3\,nm (63rd) and Fe \textsc{x} 637.4 nm (52nd) line. Line profiles in different orders were not co-added because the camera was best focused on the strongest orders, and the wavelength scale varies with orders (see more discussion in Appendix~\ref{appen:orders}). To maximize the S/N in fitting Doppler velocities and line widths, data were averaged over 5 pixels along the slit. 

\subsection{Data Analysis and Results}

\begin{figure*}[htb!]
\gridline{\fig{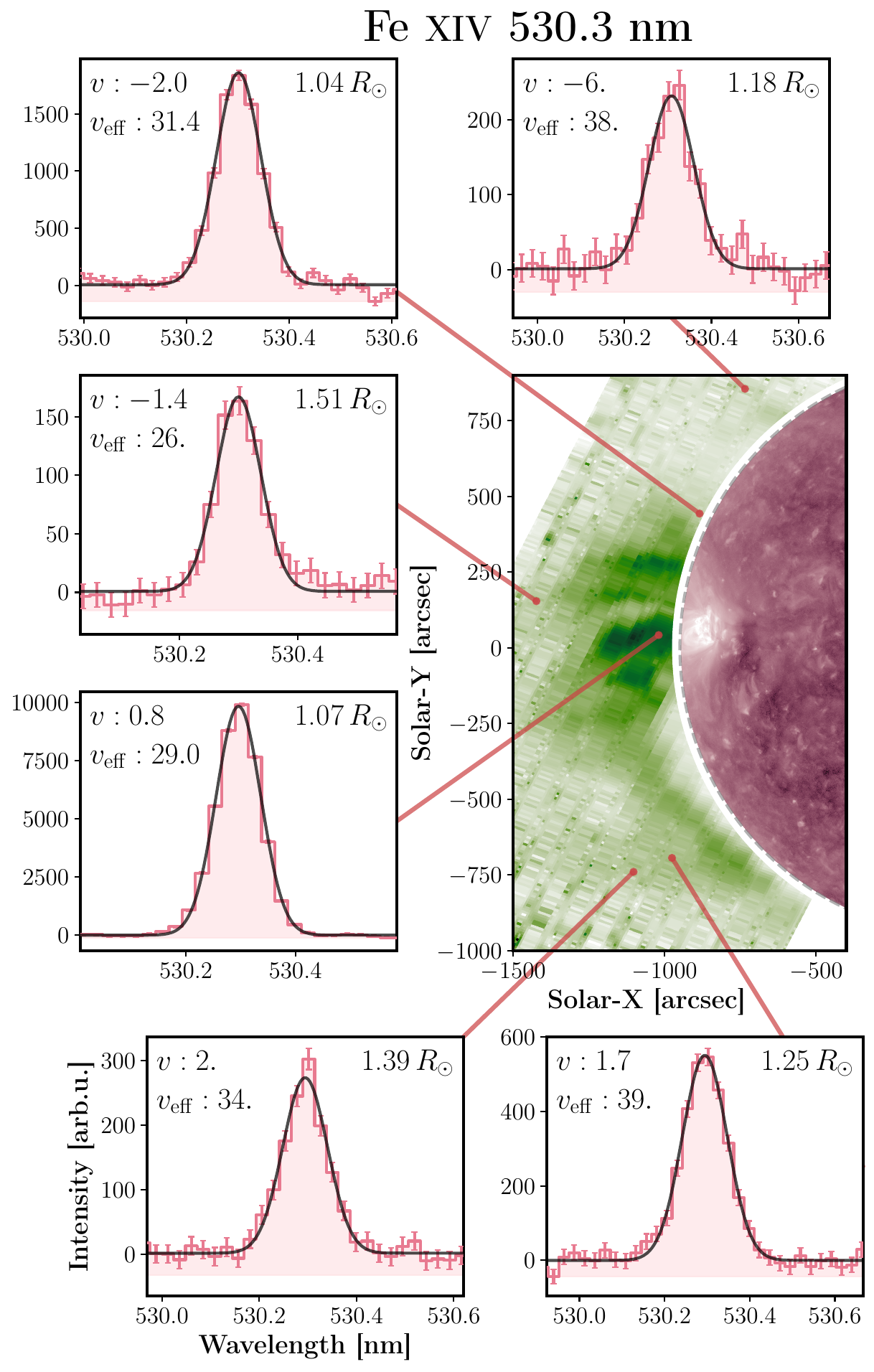}{0.44\textwidth}{(a)}
          \fig{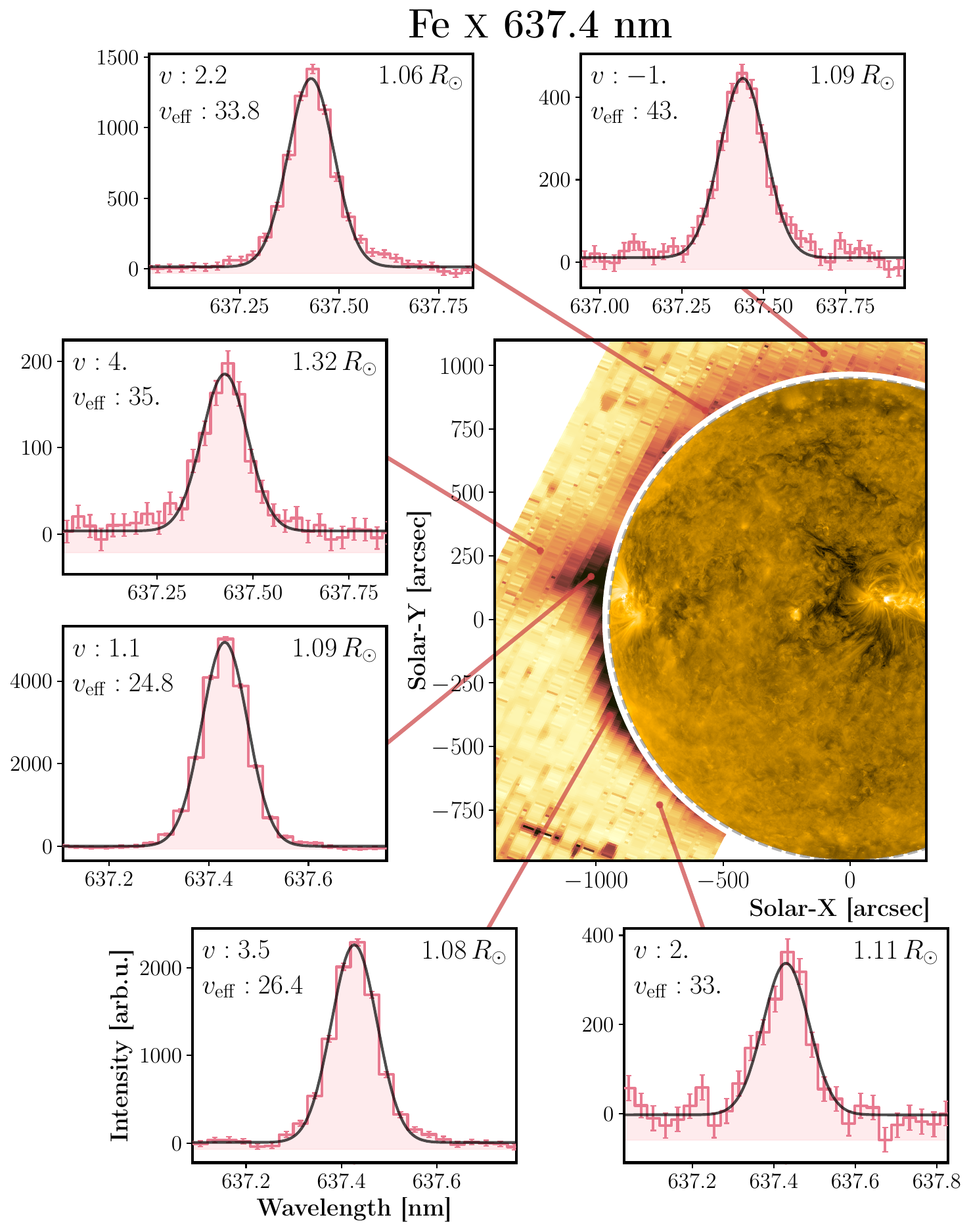}{0.55\textwidth}{(b)}}
    \caption{Overview of the Fe \textsc{xiv} 530.3\,nm (left) and Fe \textsc{x} 637.4\,nm (right) line intensities observed by 3PAMIS. The line-to-continuum ratios of the two lines are shown, along with the SDO/AIA images on the disk. To fill the data gaps, the intensity is interpolated using a 2-D Gaussian convolution kernel. The small zoom-in panels show Fe \textsc{xiv} 530.3\,nm and Fe \textsc{x} 637.4\,nm line profiles, which are binned over 5 pixels along the slit. Additionally, single-Gaussian fit results of these profiles are shown. $v$ represents the Doppler velocity, and $v_{\rm eff}$ denotes the effective velocity, both of which are in units of km\,s$^{-1}$. Link to the \texttt{Jupyter} notebook creating this figure: \href{https://yjzhu-solar.github.io/Eclipse2017/ipynb_html/off_limb_intensity_map_ext_zoomin.html}{\faGithub}.}
    \label{fig:fit_zoomin}
\end{figure*}

\begin{figure*}[htb!]
    \centering
    \includegraphics[width=\textwidth]{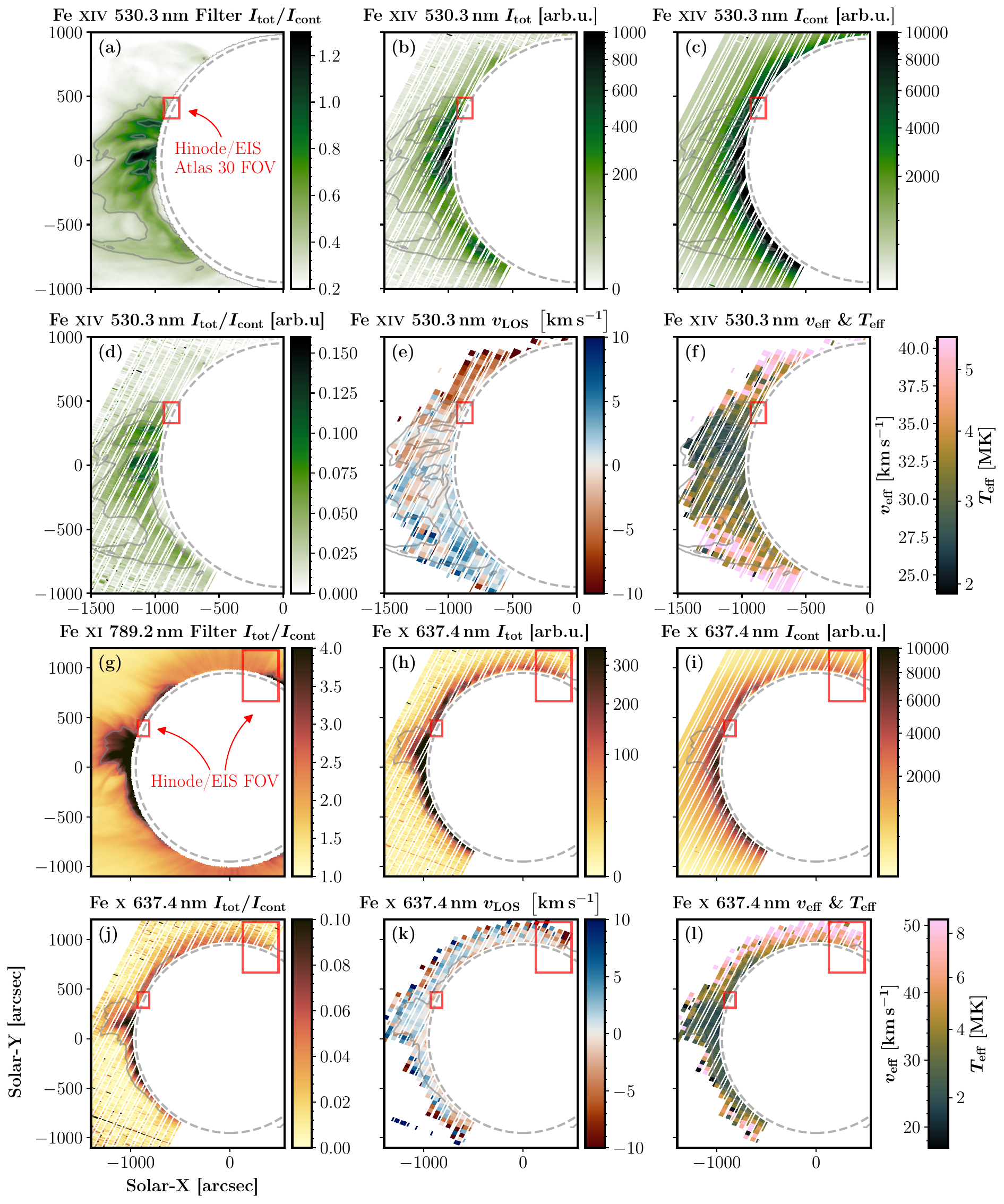}
    \caption{(a) Fe \textsc{xiv} 530.3\,nm line-to-continuum ratio measured by a narrow-bandpass imager from \citet{Boe2020}. (b) 3PAMIS Fe \textsc{xiv} 530.3\,nm line intensity. (c) ambient continuum intensity. (d) line-to-continuum ratio. (e) Doppler velocity. (f) line width. (g) Fe \textsc{xi} 789.2\,nm narrow-band image. (h-l) similar to panels (b-f) but for Fe \textsc{x} 637.4\,nm. The red rectangles represent the FOVs of two complementary Hinode/EIS observations. The gray contours outline the intensity structures in Fe \textsc{xiv} (panels a to f) and Fe \textsc{x} (panels g to l) narrow bandpass images. Link to the \texttt{Jupyter} notebook creating this figure: \href{https://yjzhu-solar.github.io/Eclipse2017/ipynb_html/off_limb_intensity_map_ms.html}{\faGithub}.}
    \label{fig:FeXIV_FeX}
\end{figure*}

Figure~\ref{fig:fit_zoomin} shows the line-to-continuum maps and line profiles of Fe \textsc{xiv} 530.3\,nm and Fe \textsc{x} 637.4\,nm observed in the off-limb regions. The profiles are obtained by averaging five pixels along the slit and removing the multi-order continuum. The intensity maps reveal several structures at the east limb, including the NOAA active region (AR) 12672 near the equator, a streamer in the northeast direction, and another streamer in the southeast \citep[also see][]{Boe2020}. Global magnetohydrodynamic (MHD) simulations and white light observations confirmed that a streamer cusp and polar plumes from a low-latitude coronal hole (CH) at the far side contributed to the emission in the northeast region \citep{Mikic2018}. With 1-3\,s exposures and spatial binning of approximately 40\arcsec along the slit, the Fe \textsc{xiv} 530.3\,nm profiles can be fitted up to 1.5\,$R_\odot$ in the AR, while the Fe \textsc{x} 637.4\,nm line can be observed up to 1.3\,$R_\odot$. 

The Fe \textsc{xiv} emission appears most prominent above the AR, while other diffuse emission is observed in the southern streamer. Notably, the Fe \textsc{xiv} 530.3\,nm profiles in the AR are narrower than the fainter Fe \textsc{xiv} profiles in streamers. Most Fe \textsc{x} emission forms close to the limb, below 1.1\,$R_\odot$, except for the northern boundary of the AR. The Fe \textsc{x} profiles in the northern CHs are remarkably broader than the Fe \textsc{x} profiles in the close-field regions. 


The fitting results of Fe \textsc{x} and Fe \textsc{xiv} are shown in Figure~\ref{fig:FeXIV_FeX}. Panel d displays the relative line-to-continuum ratio of Fe \textsc{xiv}, demonstrating similarities to the line-to-continuum ratio measured using narrow-bandpass filters shown in Figure~\ref{fig:FeXIV_FeX}a, using the technique described by \cite{Boe2020}.

Figure~\ref{fig:FeXIV_FeX}e shows the Doppler shifts in Fe \textsc{xiv} 530.3\,nm line. The Doppler velocities in the AR vary within $\pm2\,\mathrm{km\,s^{-1}}$. Larger Doppler shifts are evident in the southern and northern streamers. The northern streamer exhibits a redshift of up to 5--10\,km\,s$^{-1}$, while the southern streamer reveals a blueshift of approximately 5\,km\,s$^{-1}$.

It is often assumed that the observed full width at half maximum (FWMH) $\Delta \lambda_{\rm true}$ consists of a thermal width associated with the ion temperature $T_i$ and a non-thermal width $\xi$ caused by other unresolved motions \citep{DelZanna2018b}:
\begin{equation}
    \Delta \lambda_{\rm true} = \left[4\ln 2 \left(\frac{\lambda_0}{c} \right)^2 \left(\frac{2k_B T_i}{m_i} + \xi^2 \right)\right]^{1/2}
\end{equation}
where $\lambda_0$ is the wavelength of the spectral line, $c$ is the speed of light, $k_B$ is the Boltzmann constant, and $m_i$ represents the ion mass. To represent the width of different spectral lines with various $\lambda_0$, we introduced the effective velocity $v_{\rm eff}$ or effective temperature $T_{\rm eff}$ as
\begin{equation}
    v_{\rm eff}^2 = \frac{2k_B T_{\rm eff}}{m_i} \equiv \frac{2k_B T_i}{m_i} + \xi^2  
\end{equation}
Notably, $v_{\rm eff}$ is equivalent to the $1/e$ velocity $v_{1/e}$ used in other publications \citep[e.g.,][]{Wilhelm2005}. 
As depicted in Figure~\ref{fig:FeXIV_FeX}f, the typical Fe \textsc{xiv} line widths in the off-limb AR range from 25 to 32\,km\,s$^{-1}$. These line widths correspond to effective temperatures of 2.1--3.5\,MK. In comparison, the Fe \textsc{xiv} line profiles in the northern and southern streamers are much broader. The effective velocities in these streamers are in the range of $v_{\rm eff} = 35$--40\,km\,s$^{-1}$, equivalent to $T_{\rm eff} = 4.2$--5.4\,MK.  

Panels h to l of Figure~\ref{fig:FeXIV_FeX} display the fitting results of the Fe \textsc{x} 637.4\,nm line. Due to the lack of calibrated Fe \textsc{x} 637.4\,nm narrow bandpass images, the line intensity to continuum ratio of the Fe \textsc{xi} 789.2\,nm line, which has a similar formation temperature, is shown in panel a for comparison.  Most Fe \textsc{x} line emission forms near the limb. The enhancement of Fe \textsc{x} emission in the AR appears to be located between the two other loop-like structures observed in Fe \textsc{xiv}, indicating the existence of cold ($\sim$1\,MK) loops or plasma outflows in the vicinity of the hot (2\,MK) AR \citep[also see][]{Boe2020}.

The Doppler shifts of Fe \textsc{x} 637.4\,nm, shown in Figure~\ref{fig:FeXIV_FeX}k, range between -5 and 5\,km\,s$^{-1}$, which is within the uncertainty of the absolute wavelength calibration of 3PAMIS. In the AR and quiet Sun (QS) corona, Fe \textsc{x} shows an 
effective velocity $v_{\rm eff}$ of 20--25\,km\,s$^{-1}$, corresponding to an effective temperature $T_{\rm eff}$ of approximately 2\,MK. On the other hand, the Fe \textsc{x} line in the CH, is similar to other observations \citep[e.g.,][]{Hahn2012}, as it shows extreme broadening with  $v_{\rm eff} > 40$\,km\,s$^{-1}$. The effective temperature $T_{\rm eff}$ of Fe \textsc{x} in the CH exceeds 6\,MK, suggesting the presence of significant nonthermal velocities or additional heating of the Fe \textsc{x} ion.  

\begin{figure*}[htb]
    \centering
    \includegraphics[width=\linewidth]{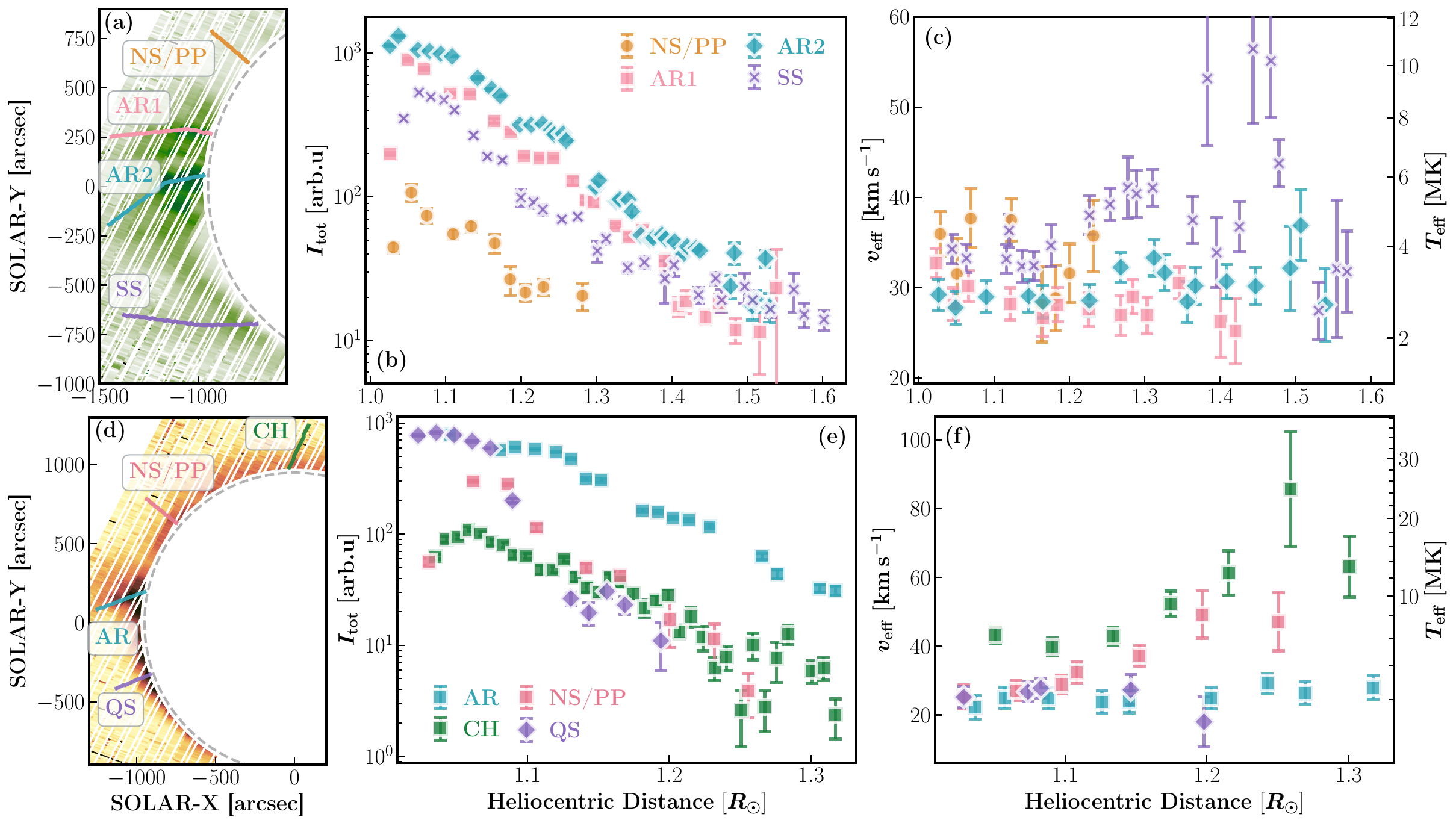}
    \caption{Fe \textsc{xiv} 530.3\,nm line intensity $I_{\rm tot}$ (b) and line width (c) variation along four cuts: northern streamer and polar plumes (NS/PP), active region 1 (AR1), active region 2 (AR2), and southern streamer (SS). Fe \textsc{x} 637.4\,nm line intensity $I_{\rm tot}$ (e) and line width (f) variation along four cuts: northern streamer and polar plumes (NS/PP), the coronal hole (CH), the active region  (AR), and the quiet Sun (QS). Link to the \texttt{Jupyter} notebook creating this figure: \href{https://yjzhu-solar.github.io/Eclipse2017/ipynb_html/off_limb_intensity_map_ms.html}{\faGithub}.}
    \label{fig:FeXIV_FeX_cut}
\end{figure*}

Panels a--c of Figure~\ref{fig:FeXIV_FeX_cut} depict the variation of Fe \textsc{xiv} 530.3\,nm line intensity and widths along four different cuts in various structures. Compared to Fe \textsc{xiv} in the AR, the Fe \textsc{xiv} lines in streamers are dimmer by at least a factor of 2. Fe \textsc{xiv} line intensities show a nearly exponential decrease in all four regions below a heliocentric distance of approximately 1.4\,$R_\odot$. Above this distance, the Fe \textsc{xiv} intensity decreases more slowly with height. This behavior might be attributed to the increase in photoexcitation to populate the upper energy level, the limitation of the instrument sensitivity, or the radial dependence in the hydrostatic and isothermal atmosphere \citep{Aschwanden2005} as
\begin{equation}
    p(r) = p_0 \exp\left[- \frac{r - R_\odot}{\lambda_{p}(T_e)}\cdot \frac{R_\odot}{r} \right]
\end{equation}
where $p_0$ is the pressure at the surface, $r$ denotes the heliocentric distance, and $\lambda_p(T_e)$ is the hydrostatic scale height. The intensity drop was fitted with the hydrostatic model along the cut in the southern streamer, noting that $I(r) \propto p(r)^2$ when collisional excitation dominates (also see Discussion in Section~\ref{dis:photoex_scatter}). We found a hydrostatic scale height $\lambda_p$ of approximately 100\,Mm in the streamer, corresponding to a temperature of around 2\,MK. 

Regarding the Fe \textsc{xiv} line widths, they appear to be nearly constant along the two cuts in the AR, showing an effective temperature $T_{\rm eff}$ of approximately 2.5 - 3 MK.

The variation of Fe \textsc{x} 637.4\, nm line intensity and width along four different cuts is shown in Panels d--f of Figure~\ref{fig:FeXIV_FeX_cut}. The Fe \textsc{x} line intensity decreases exponentially with height up to 1.3\,$R_\odot$ along these cuts. Similar to Fe \textsc{xiv}, the Fe \textsc{x} line width in the AR slightly increases from 20 to 25\,km\,s$^{-1}$ between 1.0--1.3\,$R_\odot$. The Fe \textsc{x} widths in the QS below 1.15\,$R_\odot$ also remains relatively constant at $v_{\rm eff} \approx 25$\,km\,s$^{-1}$. In contrast, the Fe \textsc{x} line widths in the CH show a drastic increase from 40 to 60\,km\,s$^{-1}$ between 1.1 and 1.2\,$R_\odot$, reaching an effective temperature greater than 10\,MK above 1.2\,$R_\odot$. Notably, the Fe \textsc{x} line widths in the northeast region appear to be slightly broader than those in the AR below 1.1\,$R_\odot$, but narrower than Fe \textsc{xiv} in the same region. However, its width increases from 25 to 50\,km\,s$^{-1}$ between 1.1--1.2\,$R_\odot$, approaching the Fe \textsc{x} widths in the CH observed by 3PAMIS 
. This behavior might be attributed to the transition of the emission from the streamer cusp to polar plumes as height increases.

\section{Comparison with Ancillary Space-based and Ground-based Data}\label{sec:space_ground}

\subsection{Hinode/EIS} \label{subsec:method_EIS}
\subsubsection{Overview and Data Reduction}

Although Hinode/EIS did not acquire observations during totality, we compared the 3PAMIS observations with two EIS observations from August 21. One observation (\texttt{dhb\_polar\_scan}) was a raster scan of the north pole CH region, while the other (\texttt{Atlas\_30}) was a QS observation with a limited FOV at the east limb as shown in Figure~\ref{fig:obs_summray}.

The first CH observation was conducted from  11:08:18 to 14:44:03 UT using the 2\arcsec\ slit. EIS made a 180-step raster scan with a step size of 2\arcsec, resulting in a FOV of 360\arcsec$\times$512\arcsec. The exposure time of each raster was 70\,s. The center of the Fe \textsc{xii} 19.51\,nm FOV was 308.4\arcsec, 920.9\arcsec in the helioprojective Cartesian coordinate.   

In the second observation, EIS made a 60-step raster scan of an off-limb QS region north of NOAA AR 12672 from 20:54:39 to UT 21:25:59 UT. Only 160\arcsec\ along the 2\arcsec\ slit were used with a step size of 2\arcsec, yielding a FOV of 120\arcsec$\times$160\arcsec. In each raster, EIS made a full CCD exposure with an exposure time of 30\,s. The center pointing of EIS at Fe \textsc{xii} 19.51\,nm was (-872.9\arcsec, 391.9\arcsec). 

The EIS level-1 HDF5 files were downloaded from the Naval Research Lab (NRL) website\footnote{\url{https://eis.nrl.navy.mil/}} and first processed using the EIS Python Analysis Code (EISPAC)\footnote{\url{https://github.com/USNavalResearchLaboratory/eispac}}. The EIS pointing was corrected by comparing the Fe \textsc{xii} 19.51\,nm intensity with the 19.3 nm broadband images taken by the Atmospheric Imaging Assembly \citep[AIA;][]{Lemen2012} on board the Solar Dynamics Observatory \citep[SDO;][]{Pesnell2012}.

Both EIS observations experienced data losses and hot pixels caused by the South Atlantic Anomaly (SAA), particularly in the QS observation. In addition, the off-limb signal-to-noise ratio (S/N) was too low to make a convincing fit of the spectral line widths. Therefore, additional data binning was required to increase the S/N. Since EISPAC does not support data binning along the slit, we developed our own codes to correct the slit tilt, average the data, and fit the spectral lines.

\subsection{Data Analysis and Results}

\begin{figure}[htb!]
    \centering
    \includegraphics[width=\linewidth]{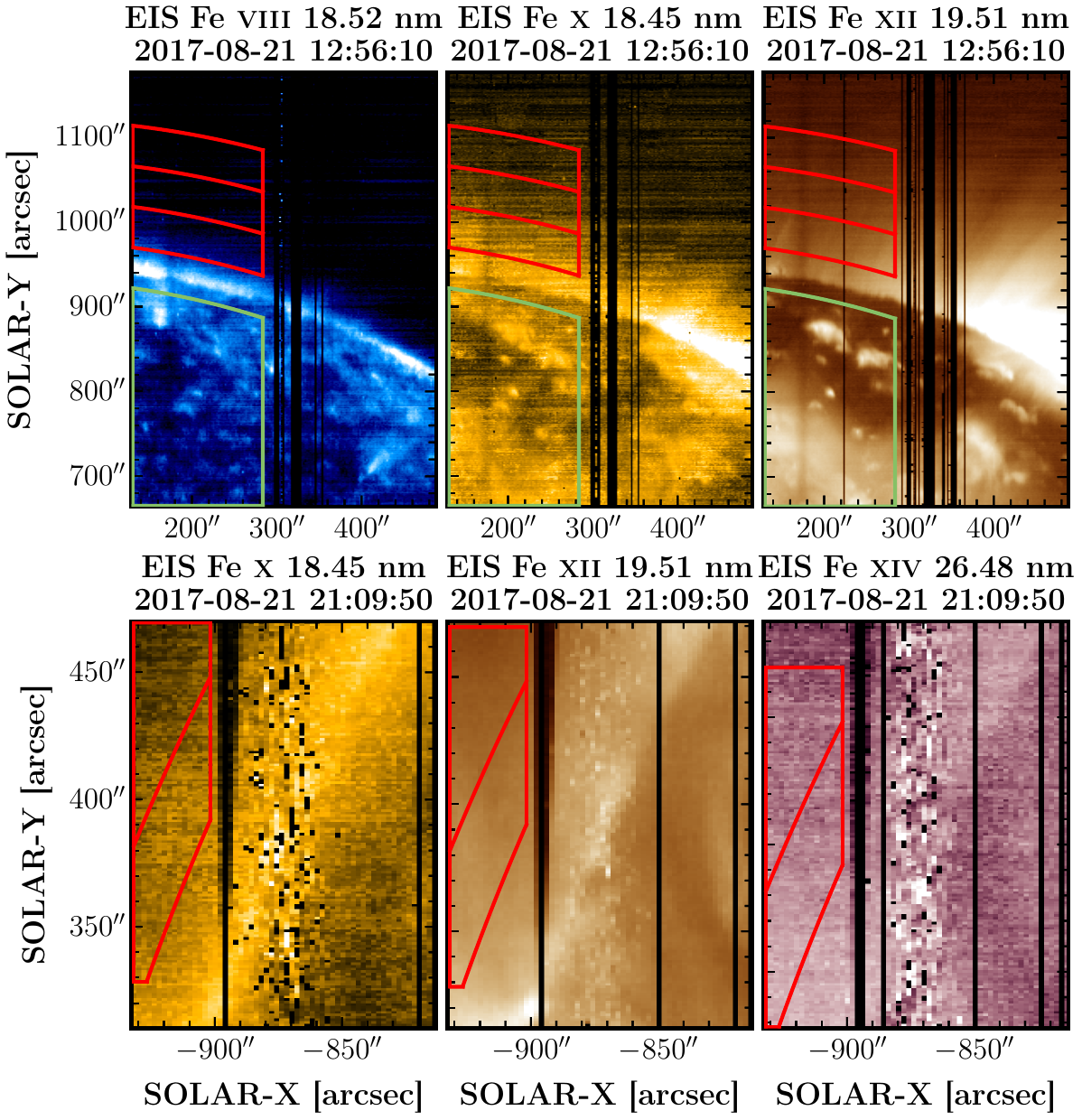}
    \caption{Overview of EIS line intensity fitted by EISPAC. Top: \texttt{dhb\_polar\_scan} observation of the CH. Bottom: \texttt{Atlas\_30} observation of the east limb QS region. The red curves highlight the regions where line profiles are averaged. Profiles in the green boxes are averaged to estimate the off-limb stray light for the CH observation. Link to the \texttt{Jupyter} notebook creating this figure: \href{https://yjzhu-solar.github.io/Eclipse2017/ipynb_html/eis_chqs_quicklook.html}{\faGithub}.}
    \label{fig:eis_quicklook}
\end{figure}

Figure~\ref{fig:eis_quicklook} summarizes the EISPAC fitting results for several prominent spectral lines observed in the CH and AR data sets. The Fe \textsc{xii} 19.5\,nm intensity map clearly outlines the boundary of the CH on the disk. Due to the data loss near the center of the FOV and the low S/N in the off-limb CH region, the EIS line profiles were averaged from three distinct regions between 1.03--1.08\,$R_\odot$, 1.08--1.13\,$R_\odot$, and 1.13--1.18\,$R_\odot$ on the left side of the EIS FOV. Additionally, the 2\% on-disk intensity was used to estimate and remove the stray light in the off-limb CH \citep{UgarteUrra2010}. 

In the QS data set, data loss and the hot pixels caused by SAA were identified at the center of FOV. Similarly, two regions were chosen, one spanning 1.035 to 1.06\,$R_\odot$ and the other between 1.06 and 1.1\,$R_\odot$, to average the EIS data. As the QS region is located close to the limb where the stray light intensity is negligible, and only a few rasters recorded the uncontaminated on-disk spectrum, no stray light correction was applied to the QS data.

\begin{figure}[htb!] 
    \centering
    \includegraphics[width=\linewidth]{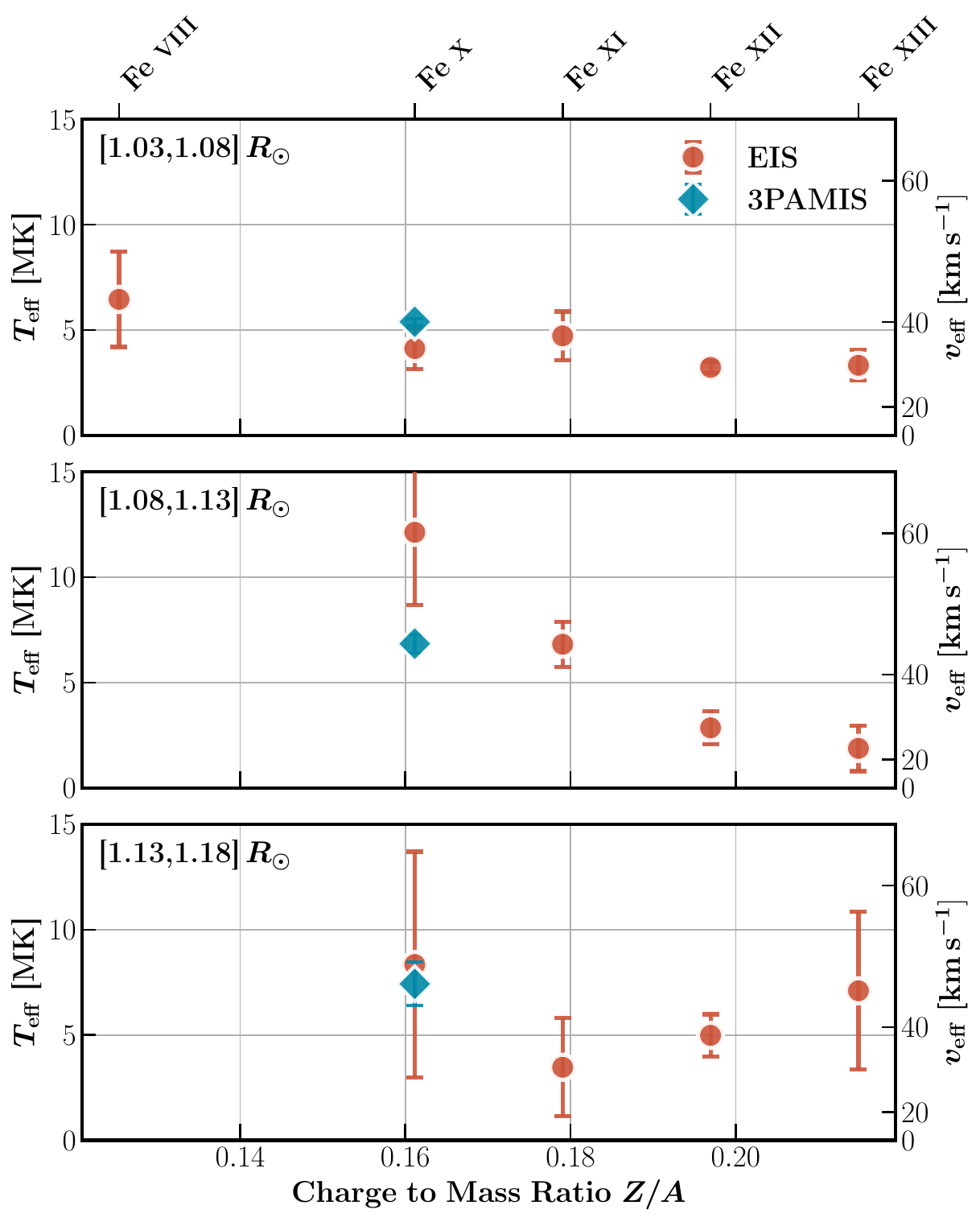}
    \caption{Effective temperature $T_{\rm eff}$ and effective velocity $v_{\rm eff}$ of different ions observed by 3PAMIS and EIS in the north pole coronal hole at three different heights: 1.03--1.08\,$R_\odot$ (top), 1.08--1.13\,$R_\odot$ (mid), and 1.13--1.18\,$R_\odot$ (bottom). Link to the \texttt{Jupyter} notebook creating this figure: \href{https://yjzhu-solar.github.io/Eclipse2017/ipynb_html/npchdb_pamis_teff.html}{\faGithub}.}
    \label{fig:npchdb_teff}
\end{figure}

Figure~\ref{fig:npchdb_teff} illustrates the relationship between $T_{\rm eff}$ and ion charge-to-mass ratio $Z/A$ of different Fe charge states observed by EIS and 3PAMIS in the off-limb CH. To measure $T_{\rm eff}$, the strongest and unblended lines were selected, including the Fe \textsc{viii} 18.52\,nm, Fe \textsc{x} 18.4\,nm, Fe \textsc{xi} 18.82\,nm, Fe \textsc{xii} 19.35\,nm, and Fe \textsc{xiii} 20.20\,nm lines. The coolest Fe \textsc{viii} 18.52\,nm line was only fitted between 1.03--1.08\,$R_\odot$ due to S/N limitations. Furthermore, the Fe \textsc{x} 637.4\,nm line widths in the same regions are averaged for comparison. 

The dependence of $T_{\rm eff}$ on ion $Z/A$, as observed by EIS, varies at different heights. Between 1.03--1.08\,$R_\odot$, Fe \textsc{viii}, which has the lowest $Z/A$, shows the highest $T_{\rm eff} \approx 6$\,MK. The $T_{\rm eff}$ of the other ions gradually decreases from 4\,MK to 3\,MK as $Z/A$ increases from 0.16 to 0.22. The decrease in $T_{\rm eff}$ for ions with $0.16<Z/A<0.22$ becomes more prominent at 1.08--1.13\,$R_\odot$, ranging from more than 6\,MK to 2\,MK. At 1.13--1.18\,$R_\odot$, $T_{\rm eff}$ first drops from about 7.5\,MK to 3\,MK at $Z/A\sim0.18$, then gradually increases to 7\,MK at $Z/A\sim0.22$. 

The Fe \textsc{x} 637.4\,nm line width observed by 3PAMIS agrees with the Fe \textsc{x} 18.4\,nm line width observed by EIS at 1.03--1.08 and 1.13--1.18\,$R_\odot$. Nevertheless, between 1.08 and 1.13\,$R_\odot$, the Fe \textsc{x} 18.4\,nm line appears to be much broader than the Fe \textsc{x} 637.4\,nm line observed by 3PAMIS, which might be caused by the low S/N and hot pixels in EIS data set. Additionally, it should be noted that the averaging of line profiles in different rasters may include additional orbital drifts not removed by the EIS software \citep{EISNote5}.

\begin{figure}[htb!]
    \centering
    \includegraphics[width=\linewidth]{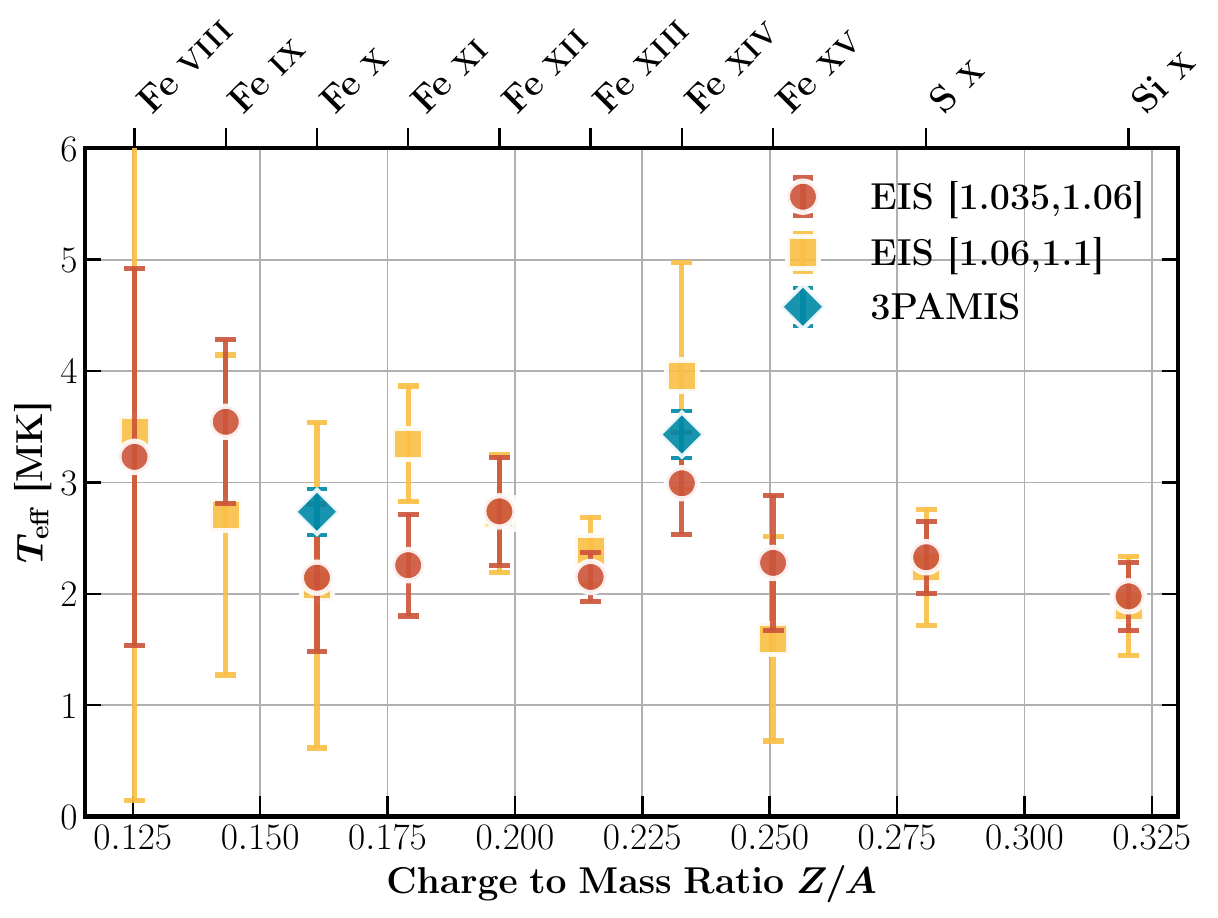}
    \caption{Effective temperature $T_{\rm eff}$ of different ions observed by 3PAMIS and EIS in the off-limb QS region. Link to the \texttt{Jupyter} notebook creating this figure: \href{https://yjzhu-solar.github.io/Eclipse2017/ipynb_html/atlas30_pamis_teff.html}{\faGithub}.}
    \label{fig:atlas30_teff}
\end{figure}

Figure~\ref{fig:atlas30_teff} shows the $T_{\rm eff}$ measured by EIS in the QS region, along with a comparison with 3PAMIS results. Benefitting from the full CCD readout and higher S/N, more EIS lines from different ions are utilized, including the Fe \textsc{ix} 19.7\,nm, Fe \textsc{xiv} 26.4\,nm, Fe \textsc{xv} 28.4\,nm, S \textsc{x} 26.4\,nm, and Si \textsc{x} 25.8\,nm lines. The EIS line widths in the two regions exhibit similarities, while $T_{\rm eff}$ at 1.06--1.1\,$R_\odot$ shows a broader distribution, likely due to the lower S/N. 

Between 1.035 and 1.06\,$R_\odot$, $T_{\rm eff}$ for most ions observed by EIS range from 2--3\,MK. Notably, the Fe \textsc{viii} and Fe \textsc{ix} with the lowest $Z/A$ display slightly higher temperatures compared to the other ions. However, $T_{\rm eff}$ does not show a distinct variation with respect to $Z/A$ in the QS data set, which precludes support for the hypothesis of preferential heating of heavy ions in the QS region at 1.035--1.1\,$R_\odot$. Yet, the $T_{\rm eff}$ values for of Fe \textsc{x} and Fe \textsc{xiv}, measured independently by EIS in EUV and 3PAMIS in the visible, are consistent with each other.

EIS observations provide additional plasma diagnostics, including the electron density $n_e$ and electron temperature $T_e$. Table~\ref{tab:atlas30_neTe} presents the measured $n_e$ and $T_e$ for the two QS regions, employing three different onboard radiometric corrections reported by \citet[][,GDZ]{DelZanna2013}, \citet[][HPW]{Warren2014}, and the latest \citet{DelZanna2023b}. The Fe \textsc{xi} and Fe \textsc{xii} diagnostics results using the latest radiometric corrections are quite different from the other two, which is probably due to the correction of the wavelength-dependent degradation of the detectors after 2012. Notably, the new EIS radiometric calibration is still under review, so we focused on the results using the first two corrections. Besides, density diagnostics using Si \textsc{x} ratios between 1.06 and 1.1\,$R_\odot$ show great uncertainties due to the low S/N.

Both regions show typical QS $n_e$ and $T_e$ 
\citep[e.g.,][]{Laming1997,Kamio2012,Feldman1999,Brooks2009}. In Fe \textsc{xii}, $n_e$ drops from $2.0\times 10^8$\,cm$^{-3}$ to $1.5\times 10^8$\,cm$^{-3}$, while $T_e$ increases from 1.2\,MK to 1.5\,MK in Fe \textsc{xi}. The HPW and GDZ methods yield similar diagnostic results for Fe \textsc{xii}. However, the HPW method provides a higher $T_e$ using Fe \textsc{xi} than the GDZ method. Additionally, $n_e$ inferred from the Si \textsc{x} 25.8/26.1 ratio is much lower if the HPW method is used, amounting to about 50--60\% of the GDZ values. 

Consistent with numerous other observations \citep[e.g.,][]{Hassler1990,Seely1997,Banerjee1998}, $T_{\rm eff}$ derived from the line widths were found to be higher than $T_e$. This implies the existence of unresolved nonthermal motions in both QS and CHs. Assuming $T_i \approx T_e$ in the QS region, we estimate a nonthermal velocity of approximately 15--25\,km\,s$^{-1}$ for ions with $T_{\rm eff}$=2--3\,MK. Additionally, the $Z/A$ dependence of $T_{\rm eff}$ in the CH suggests preferential heating of heavy ions at the base of polar CHs.

\begin{table}[htb!]
\centering
\begin{tabular}{@{}cccccc@{}}
\toprule
\toprule
\multicolumn{6}{c}{$n_e$ ($10^8$\,cm$^{-3}$)}                                                  \\ \midrule
Ion             & Line              & Region & GDZ      & HPW  & New                  \\ \midrule
Si \textsc{x}   & 25.8/26.1         & 1      & $2.00^{+0.57}_{-0.55}$ & $1.35^{+0.47}_{-0.44}$  & $2.00^{+0.63}_{-0.49}$ \\
Si \textsc{x}   & 25.8/26.1         & 2      & $0.91^{+0.56}_{-0.49}$ & $\mathbf{0.45^{+0.91}_{-0.45}}$ & $1.05^{+0.53}_{-0.50}$     \\
Fe \textsc{xii} & 18.6/19.3     & 1      & $2.09^{+0.10}_{-0.09}$ & $2.00^{+0.09}_{-0.09}$ & $1.45^{+0.03}_{-0.07}$\\
Fe \textsc{xii} & 18.6/19.3     & 2      & $1.58^{+0.16}_{-0.10}$ & $1.51^{+0.15}_{-0.10}$ &  $1.12^{+0.08}_{-0.10}$  \\ \midrule
\multicolumn{6}{c}{$T_e$ ($10^6$\,K)}                                                          \\ \midrule
Fe \textsc{xi}  & 18.8/25.7 & 1      & $1.20^{+0.15}_{-0.13}$ & $1.51^{+0.19}_{-0.16}$ & $2.19^{+0.26}_{-0.28}$\\
Fe \textsc{xi}  & 18.8/25.7 & 2      & $1.48^{+0.43}_{-0.38}$ & $1.86^{+0.59}_{-0.54}$ & $>1.95$\\ \bottomrule
\end{tabular}
\caption{Electron density $n_e$ and electron temperature $T_e$ diagnostics of regions 1 (1.035--1.06\,$R_\odot$) and 2 (1.06--1.1\,$R_\odot$). Fe \textsc{xii} 18.68\,nm and Fe \textsc{xi} 25.75\, nm lines are self-blended. Entries in the GDZ, HPW, and New columns utilize the radiometric corrections reported by \citet{DelZanna2013}, \citet{Warren2014}, and the latest \citet{DelZanna2023b}, respectively.}
\label{tab:atlas30_neTe}
\end{table}

\subsection{CoMP}

CoMP is a tunable coronagraph located at the Mauna Loa Solar Observatory (MLSO). CoMP can perform spectropolarimetric observations of Fe \textsc{xiii} 1074.7 and 1079.8\,nm lines in the near-infrared between 1.05--1.35\,$R_\odot$. The Stokes parameters $I$, $Q$, $U$, and $V$ are sampled at 3 or 5 wavelength positions across the Fe \textsc{xiii} profiles using Lyot filters. The three-point Stokes $I$ profiles are inverted analytically to obtain the line intensity, Doppler shifts, and widths \citep{Tian2013}. During the 2017 August 21 TSE, CoMP carried out observations from 17:05 to 18:19 UT. We utilized the median Doppler velocity at the east limb as the zero point velocity. An instrumental width of 21\,km\,s$^{-1}$ was removed during the data reduction \citep{Morton2015}. 

\begin{figure*}[htb]
    \centering
    \includegraphics[width=0.95\linewidth]{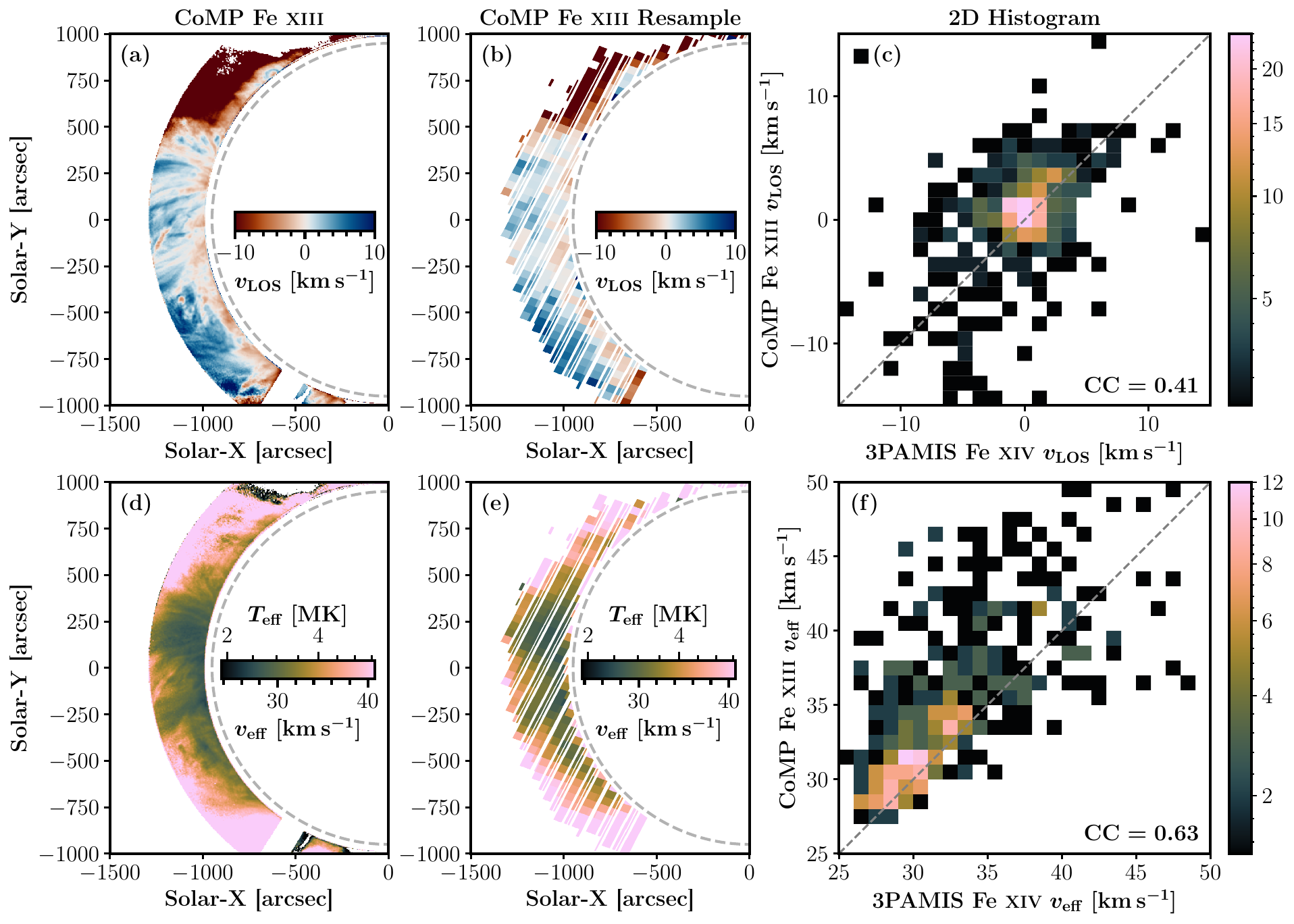}
    \caption{Comparison between the Doppler velocity $v_{\rm LOS}$ and line width observed in Fe \textsc{xiii} 1074.7\,nm by CoMP and Fe \textsc{xiv} 530.3\,nm by 3PAMIS. (a) CoMP  Doppler velocity. (b) CoMP Doppler velocity resampled to match the 3PAMIS pixels. (c) 2D histogram and Pearson correlation coefficient of the Doppler velocity measured by 3PAMIS and CoMP. (d) CoMP Fe \textsc{xiii} line widths. (e) CoMP widths resampled at 3PAMIS pixels. (f) 2D histogram and Pearson correlation coefficient of the widths measured by 3PAMIS and CoMP. Link to the \texttt{Jupyter} notebook creating this figure: \href{https://yjzhu-solar.github.io/Eclipse2017/ipynb_html/comp_pamis_comp.html}{\faGithub}.}
    \label{fig:PaMIS_CoMP}
\end{figure*}

The inverted Doppler velocities and line widths from the CoMP average file are compared to 3PAMIS observations in Figure~\ref{fig:PaMIS_CoMP}. The comparison is focused on the Fe \textsc{xiv} line widths and Doppler velocity, given the proximity of this emission line to the formation temperature of Fe \textsc{xiii} observed by CoMP. To make a fair comparison, the CoMP values were resampled at the same pixel scales as 3PAMIS by box averaging and compared with 3PAMIS values using 2D histograms.

The northern streamer is dominated by a redshift of approximately 10\,km\,s$^{-1}$, found to be in agreement with 3PAMIS observations. The equatorial AR, on the other hand, shows no significant Doppler shifts greater than 5\,km\,s$^{-1}$, which slightly differs from the tiny redshifts in 3PAMIS observations. In the southern streamers, blueshifts ranging from 5 to 10\,km\,s$^{-1}$ were found in CoMP observations, slightly greater than 3PAMIS values. In addition, some minor redshifts of less than 2\,km\,s$^{-1}$ were observed at the bottom of the FOV. The 2D histogram reveals some correlation between the Doppler velocity measured by CoMP and 3PAMIS, with a Pearson correlation coefficient of 0.41. However, no significant systematic Doppler shifts $>2$\,km\,s$^{-1}$ were found between the 3PAMIS and CoMP observations. Most differences in Doppler shifts are within the uncertainty of $5\,$km\,s$^{-1}$ in the 3PAMIS absolute wavelength calibration.

The line widths observed by CoMP and 3PAMIS also reveal a high level of agreement. The line widths obtained by CoMP are narrower in the equatorial regions, corresponding to $T_{\rm eff} \approx 3\,$MK. In contrast, broader line profiles ($T_{\rm eff}>4$\,MK) were found in the northern and southern streamers. The 2D histogram confirmed a good correlation between $v_{\rm eff}$ observed by CoMP and 3PAMIS from 25--35\,km\,s$^{-1}$, predominantly from the equatorial AR and the streamer structures in the vicinity. Notably, in the data bins with a count of more than 5, CoMP widths were found to be 1--3\,km\,s$^{-1}$ (approximately 5--10\%) greater than the widths observed by 3PAMIS.   

\subsection{Comparison with Other Observations}
\begin{figure*}[htb]
    \centering
    \includegraphics[width=\linewidth]{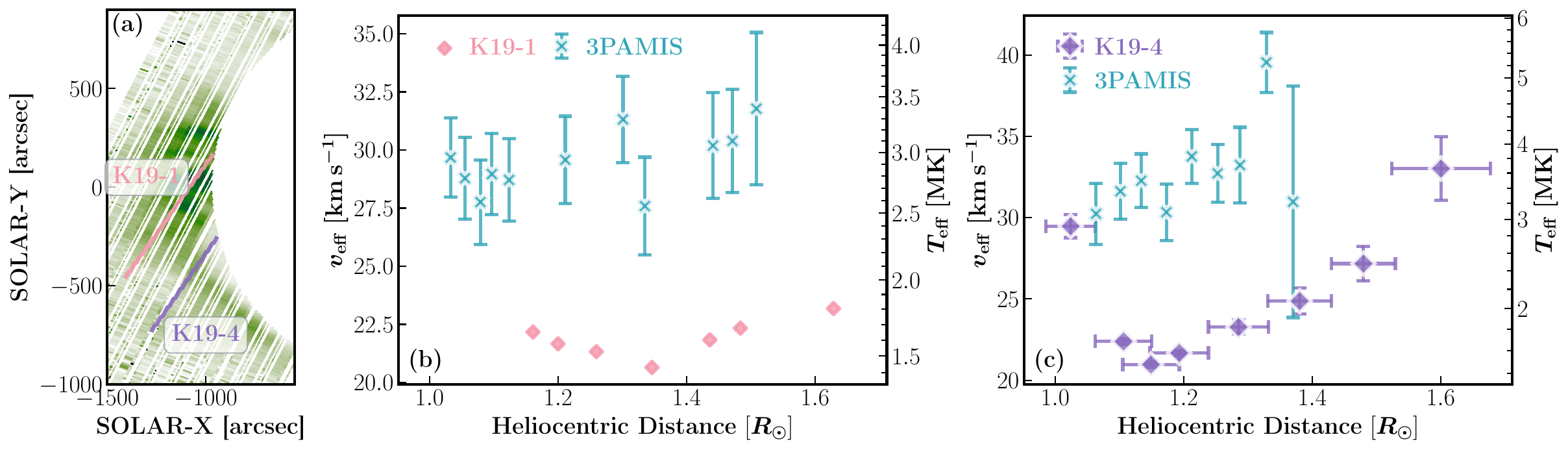}
    \caption{Comparison between the Fe \textsc{xiv} 530.3\,nm line widths given as effective velocity $v_{\rm eff}$ in (b) and effective temperature $T_{\rm eff}$ (c) measured by \citet{Koutchmy2019} (red) and 3PAMIS along the slit position 1 (K19-1) and 4 (K19-4) shown in panel (a). \citet{Koutchmy2019} (blue, purple). Link to the \texttt{Jupyter} notebook creating this figure: \href{https://yjzhu-solar.github.io/Eclipse2017/ipynb_html/off_limb_intensity_map_ms.html}{\faGithub}.}
    \label{fig:FeXIV_Teff_K19}
\end{figure*}

\citet[][hereafter K19]{Koutchmy2019} performed a slit spectroscopic experiment during the 2017 TSE and recorded coronal deep spectra from 510\,nm to 590\,nm at six different positions. Fortunately, two positions (Positions 1 and 4) were at the east limb, overlapping with the FOV of 3PAMIS. Fe \textsc{xiv} line widths along these two positions, as digitized from Figures 8 and 9 of K19, are compared with 3PAMIS observations in Figure~\ref{fig:FeXIV_Teff_K19}. Position 1 passes the AR and southern streamer, while Position 4 only covers the southern streamer.

Overall, the Fe \textsc{xiv} line widths $v_{\rm eff}$ observed by 3PAMIS were found to be approximately 40\% greater compared to those reported by K19. However, 3PAMIS and K19 revealed similar trends in the variation of line widths. Along Position 1 (AR), 3PAMIS observed a relatively constant $T_{\rm eff} \approx 3\,$MK, while K19 found a lower $T_{\rm eff} \approx 1.6\,$MK. In the southern streamer, both 3PAMIS and K19 exhibited an increase in Fe \textsc{xiv} line widths with height. In 3PAMIS observations, Fe \textsc{xiv} $T_{\rm eff}$ increased slightly from approximately 3 to 4\,MK between 1.1--1.3\,$R_\odot$. On the other hand, $T_{\rm eff}$ in K19 decreased  from 3 to 1.5\,MK between 1.0 and 1.15\,$R_\odot$, followed by a gradual increase to 3.5\,MK at 1.6\,$R_\odot$. The differences between the 3PAMIS and K19 could be attributed to the uncertainty in instrumental widths and/or uncertainty in the coalignment between the two instruments.

\section{Discussion}\label{sec:dis}
\subsection{Line Widths and Their Variation with Height: Open and Closed Fields}
The Fe \textsc{x} and Fe \textsc{xiv} line profiles observed by 3PAMIS during the 2017 TSE reveal substantial line width variations within and between different structures, especially between the open and closed field regions. In the open fields, line widths are observed to be broader and to increase with height below 1.3\,$R_\odot$, while the line widths in closed fields appear to be narrower and nearly constant.

\begin{figure*}
    \centering
    \includegraphics[width=0.9\linewidth]{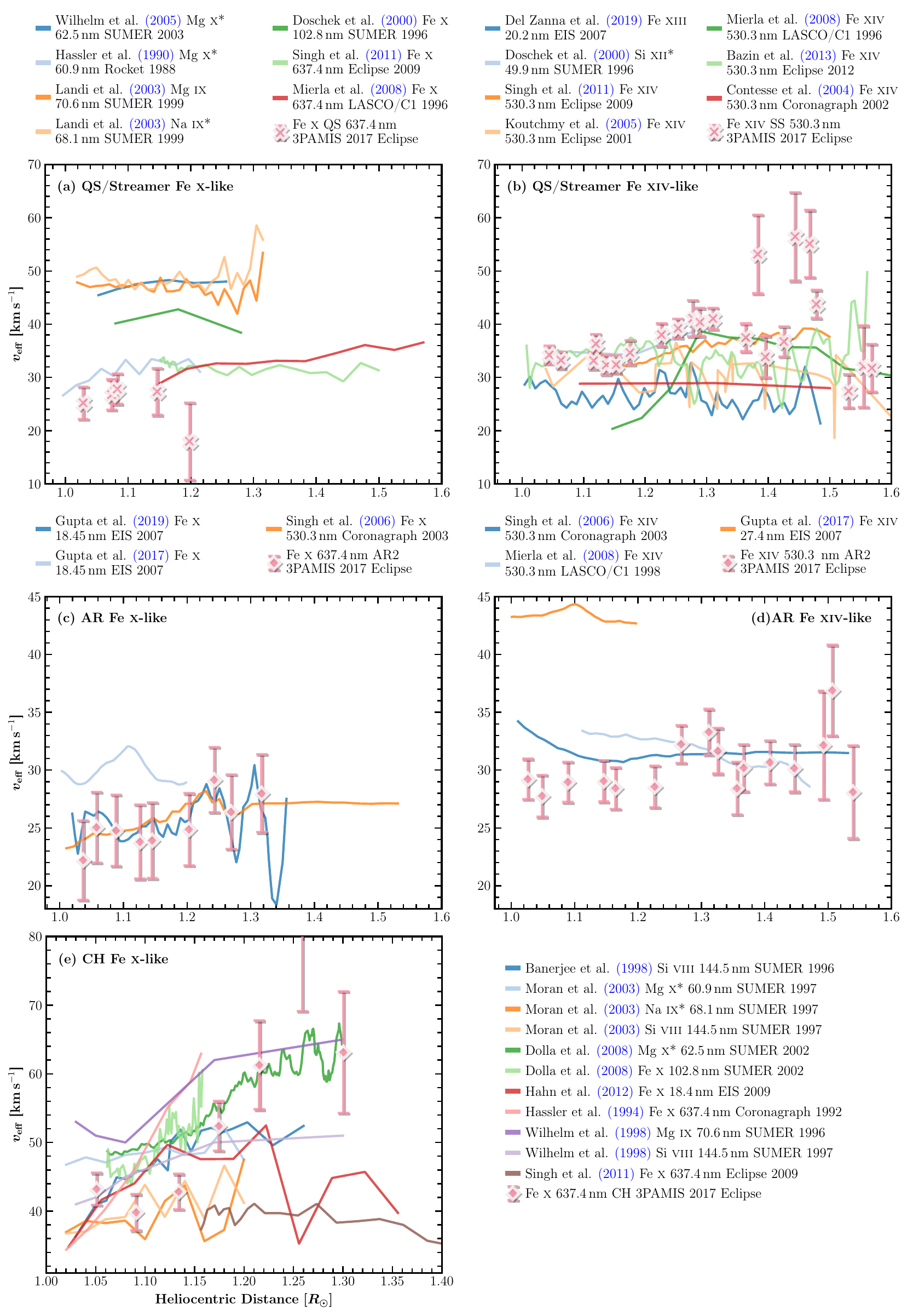}
    \caption{Comparison of the line widths variation different structures observed by 3PAMIS and other instruments, along with the spectral lines and years of observation. Lithium-like ions are labeled by *. Data is digitized from the listed publications. Link to the \texttt{Jupyter} notebook creating this figure: \href{https://yjzhu-solar.github.io/Eclipse2017/ipynb_html/huge_comparison.html}{\faGithub}. \nocite{Wilhelm2005, Hassler1990,Landi2003,Doschek2000,Singh2011,Mierla2008, DelZanna2019, Koutchmy2005, Bazin2013, Contesse2004,Gupta2019, Gupta2017, Singh2006, Banerjee1998, Moran2003, Dolla2008, Hahn2012, Hassler1994, Wilhelm1998}}
    \label{fig:huge_compare}
\end{figure*}

In Figure~\ref{fig:huge_compare}, we compared the observed line widths in different open- and closed-field regions during the TSE with the observed line widths reported by a great number of previous studies. These studies used UV or visible emission lines with similar formation temperatures to Fe \textsc{x} and Fe \textsc{xiv}. The lithium-like ions are also labeled because they usually have high-temperature tails in the equilibrium charge state population. Notably, the absolute magnitudes of $v_{\rm eff}$ are not necessarily the same because of different ion masses and plasma conditions. 

The line width variations observed by 3PAMIS are in general agreement with previous studies using UV and visible observations, which reveal minor variations in the closed fields and an increase in the open fields. For instance, a similar increase-then-decrease of Fe \textsc{xiv} widths in streamers was also reported by \citet{Mierla2008}.

To effectively address the differences in line widths across various structures or heights, we need to consider both changes in the local parameters, such as ion temperature and nonthermal velocity, and the potential influence of radiative processes on the line width. These processes may include the integration of emission with divergent Doppler shifts along the LOS, nonequilibrium ionization, photoexcitation, and resonant scattering \citep{Gilly2020}. 


We chose not to delve into the nonequilibrium ionization in this paper for two primary reasons. First, it requires comprehensive modeling of the coronal and solar wind plasma, which is beyond the scope of this work. Second, the nonequilibrium ionization does not directly affect the widths of local emissivity, while it potentially modifies the profiles through the LOS integration.  

\subsubsection{Photoexcitation and Resonant Scattering} \label{dis:photoex_scatter}
As the height increases and density decreases, photoexcitation and resonant scattering become increasingly important. The visible forbidden lines are photoexcited by the white light continuum emission from the photosphere. 
Except for the Fe \textsc{i} 530.23\,nm line at the blue wing of Fe \textsc{xiv} 530.3\,nm, the continuum has no other features and does vary significantly across the profile. Therefore, for Case I scattering between two sharp levels, the Gaussian-like photon redistribution function will convolve with the nearly flat continuum, resulting in a Gaussian emissivity profile. The width of the Gaussian emissivity is still determined by $v_{\rm eff}$, as explained in Appendix~\ref{app:scattering}. This is in contrast to the strong UV lines (e.g., Mg \textsc{x}, O \textsc{vi}, and Ly$\alpha$), as they could be photoexcited by their own profiles from the disk, affecting the width of the local emissivity profile. 

\begin{figure*}[htb!]
    \centering
    \includegraphics[width=\linewidth]{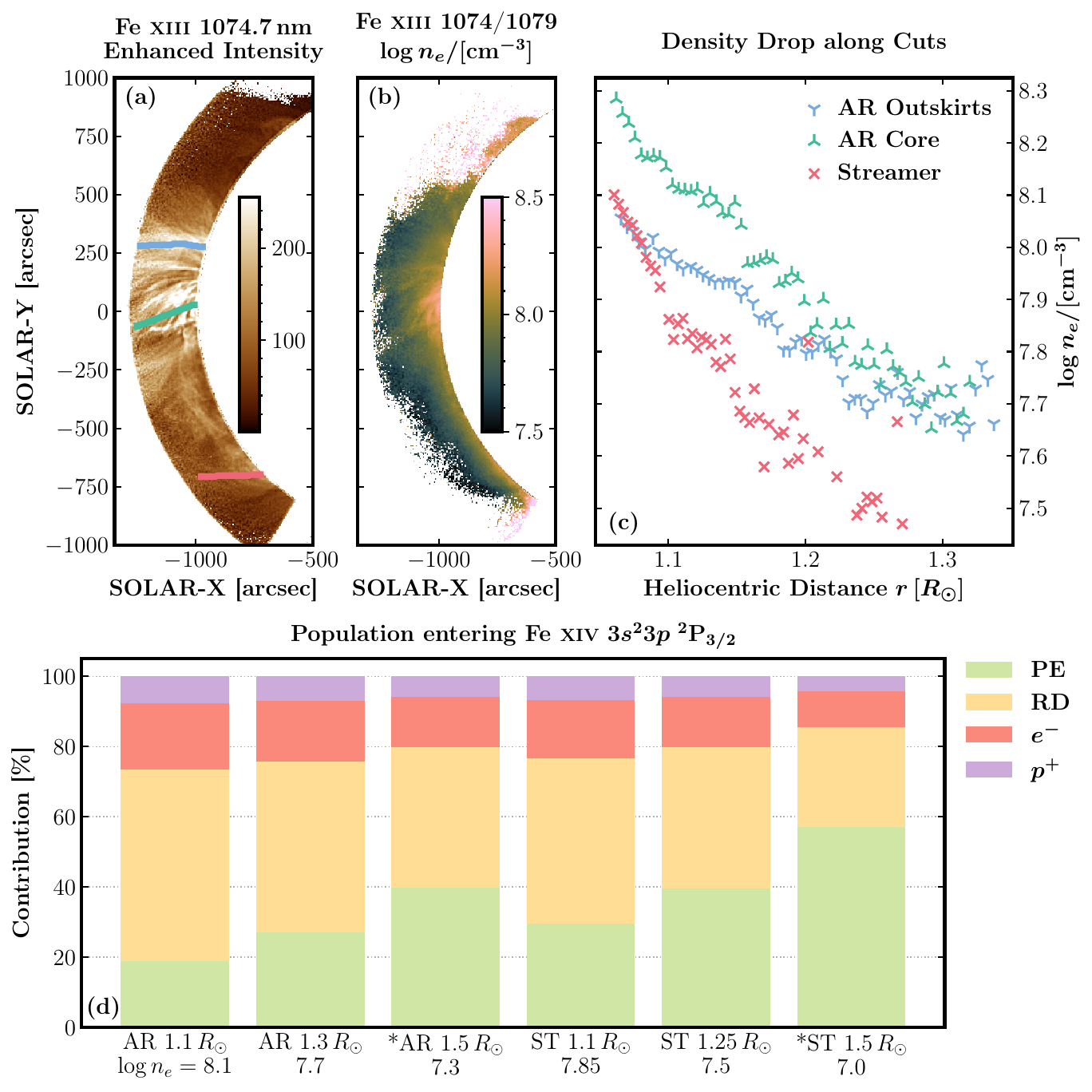}
    \caption{Electron density $n_e$ measured by CoMP and the excitation of the upper energy level of Fe \textsc{xiv}. (a) Fe \textsc{xiii} 1074.7\,nm intensity enhanced by a radial intensity filter. (b) Electron density $n_e$ inferred by Fe \textsc{xiii} 1074/1079 ratio. (c) Decrease in $n_e$ along the three cuts shown in Panel a. (d) Population entering the upper energy level of Fe \textsc{xiv} 530.3\,nm by different processes at various densities and heights. PE: photoexcitation; RD: radiative decay through cascades from higher levels; $e^-$: collisions with electrons; $p^+$: collisions with protons. *$n_e$ is extrapolated, assuming an exponential drop with height. Link to the \texttt{Jupyter} notebook creating this figure: \href{https://yjzhu-solar.github.io/Eclipse2017/ipynb_html/comp_density.html}{\faGithub}. 
    }
    \label{fig:FeXIII_dens_FeXIV_pop}
\end{figure*}

Although photoexcitation does not directly alter the widths of local emissivity, it might affect the emissivity from various structures integrated along the LOS. We estimated the contribution of photoexcitation to populate the upper energy level of Fe \textsc{xiv} in Figure~\ref{fig:FeXIII_dens_FeXIV_pop}, using the electron density inferred from Fe \textsc{xiii} 1074/1079 ratios observed by CoMP. The photoexcitation and three other collisional processes are considered, including electron and proton collisions and radiative decay from higher, collisionally-populated levels. Due to the limitation of the CoMP FOV and S/N, we extrapolated $n_e$ to 1.5\,$R_\odot$, assuming an exponential decrease.  In the AR at 1.5\,$R_\odot$, where $\log n_e \approx 7.3$, photoexcitation only contributes a maximum of 40\% of the population. In the streamer, the photoexcitation dominates the level population where $\log n_e$ drops to 7.0 at 1.5\,$R_\odot$. Therefore, we suggest that the photoexcitation and resonant scattering may neither significantly influence the line broadening, nor account for the observed discrepancies in line widths between the open and closed fields below 1.2\,$R_\odot$, in agreement with early studies by \citet{Raju1991}. Quantitative investigations on the influence of photoexcitation and resonant scattering on line broadening above 1.3 $R_
\odot$ necessitates a comparison between these observations and global MHD simulations \citep[e.g., the Alfv\'en Wave Solar Model;][]{vanderHolst2014}, which is outside the purview of the study.


\subsubsection{LOS Integration}
Emission originating from multiple structures with various macroscopic Doppler velocities or line widths can contribute to the integral along the LOS in the optically thin plasma. This effect might be evident in open-field structures where fast outflows in the lower corona are expected. Consequently, different Doppler shifts along the LOS may broaden profiles in open-field structures \citep[e.g.,][]{Akinari2007, Zhu2023}. 

On the other hand, the relatively constant line width in closed-field structures suggests that the integration of multiple structures along the LOS might not play an important role in the broadening of closed-field profiles. This implication is also supported by the fact that the $v_{\rm eff}$ of approximately 25\,km\,s$^{-1}$ measured in a single AR loop \citep{Gupta2019} is quite similar to the Fe \textsc{x} $v_{\rm eff}$ obtained by 3PAMIS. If the LOS integration is important, 3PAMIS, with its low spatial resolution, should have observed a larger excess width in the core of AR, where several coronal loops overlap along the LOS, which was not the case.  

\citet{Gilly2020} found that a relatively constant line width in the lower corona might be an illusion caused by the LOS integration below the height where the density of the ion charge state reaches its maximum. According to their polar CH and streamer model, they suggested that the Fe \textsc{x} 18.45\,nm line width appears constant below 20\,Mm, and the width of Si \textsc{xii}, which has a similar formation temperature to Fe \textsc{xiv}, remains constant below 200\,Mm. In contrast, during the eclipse, we observed constant line widths up to 200\,Mm (Fe \textsc{x}) and 350\,Mm (Fe \textsc{xiv}) in the AR. Furthermore, the off-limb AR region consists of the hottest and densest plasma near the POS, surrounded by cooler and more tenuous QS regions, which is opposite to the conditions in the polar CHs. Hence, we argue that the constant line widths cannot be solely explained by the plateauing effect proposed by \citet{Gilly2020}.

Finally, the open and close-field structures may overlap along the LOS, which makes it more challenging to interpret the behavior of line widths \citep[e.g.,][]{Zhu2021}. This often occurs at the boundary of the open and closed-field regions, such as the variation of Fe \textsc{x} widths in the northern stream and polar plumes regions and Fe \textsc{xiv} widths in the southern streamer observed by 3PAMIS  (see Figure~\ref{fig:FeXIV_FeX_cut}f).

\subsubsection{Preferential Heating}

Low $Z/A$ ions like Fe \textsc{x} are found to be preferentially heated by ion cyclotron waves in CHs, creating excessive thermal broadening \citep[e.g.,][]{Tu1998,Dolla2008,Landi2009}. In this study, the excessive heating to $Z/A$ ions, such as Fe \textsc{viii}, Fe \textsc{x}, and Fe \textsc{xi}, was also found in the EIS observation of the polar CH. Previous studies reported $T_{\rm eff}$=4--6\,MK for Fe \textsc{x} at the base of a polar CH during the solar minimum \citep{Hahn2010,Zhu2023}, which is consistent with 3PAMIS and EIS observations during the 2017 TSE. 

In principle, the constant Fe \textsc{xiv} widths in QS and streamers could potentially be dominated by thermal broadening when nonthermal motions are negligible \citep{Muro2023}. However, this scenario suggests excessive heating of both Fe \textsc{x} and Fe \textsc{xiv} with charge-to-mass ratios (0.16 and 0.23, respectively) in the QS corona.

In contrast to \citet{Muro2023}, EIS and 3PAMIS found no evidence of preferential heating in the QS plasma, in agreement with \citet{Landi2007} who studied the QS plasma during solar minimum. In ARs, frequent collisions between ions and electrons can result in $T_i \approx T_e$ \citep{Hara1999}. This is a common assumption supported by observations \citep[e.g.,][]{Imada2009} and simulations \citep[e.g.,][]{Shi2022}. 

\begin{figure*}[htb!]
    \centering
    \includegraphics[width=\linewidth]{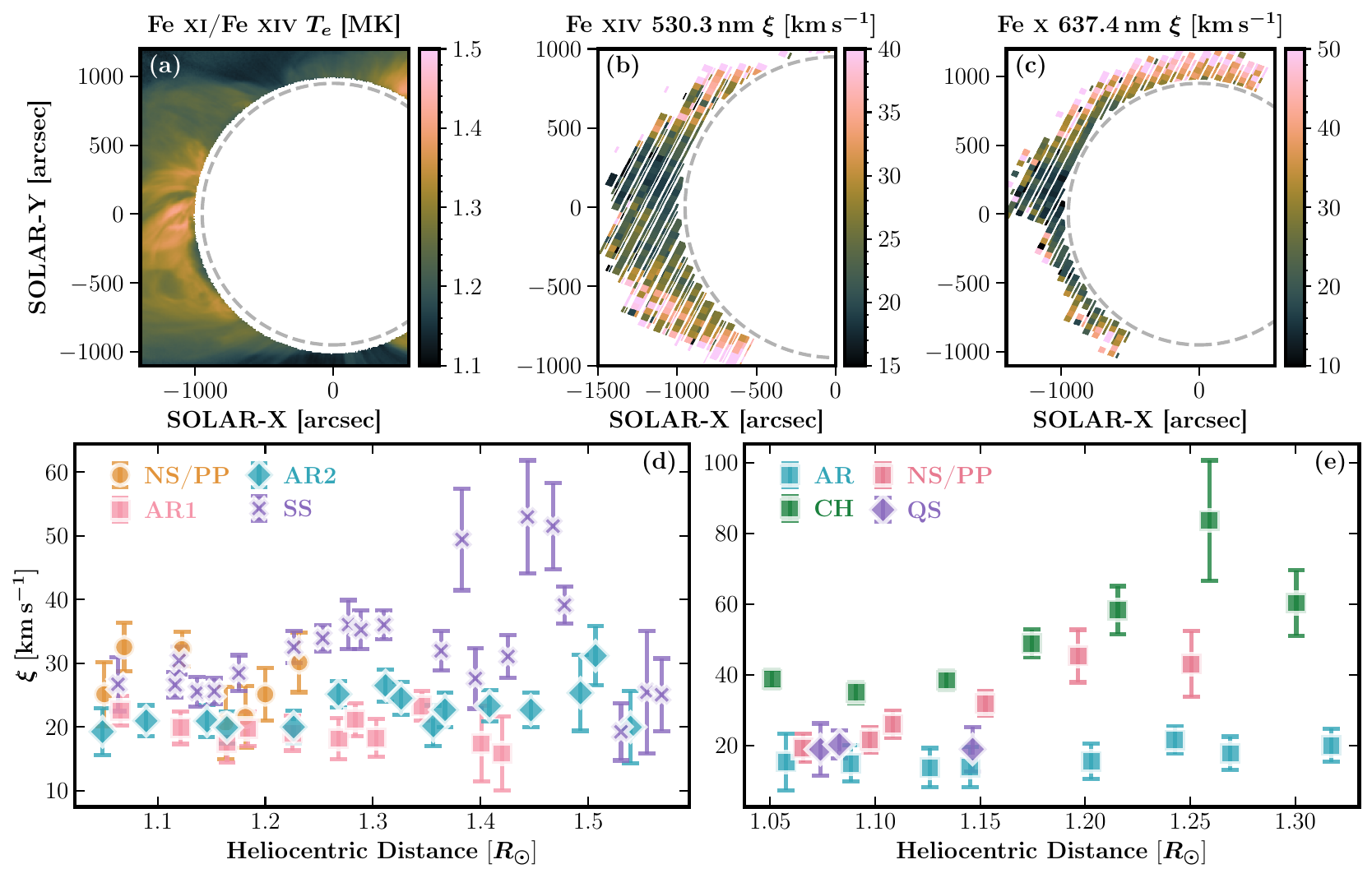}
    \caption{(a) Electron temperature $T_e$ measured by \citet{Boe2020} using the Fe \textsc{xi} 789.2\,nm and Fe \textsc{xiv} 530.3\,nm ratio. (b) and (c) Nonthermal velocity $\xi$ in Fe \textsc{xiv} 530.3\,nm and Fe \textsc{x} 637.4\,nm lines. (d) and (e) Nonthermal velocity along the cuts shown in Figure~\ref{fig:FeXIV_FeX_cut}. Link to the \texttt{Jupyter} notebook creating this figure: \href{https://yjzhu-solar.github.io/Eclipse2017/ipynb_html/off_limb_intensity_map_ms.html}{\faGithub}.}
    \label{fig:FeX_FeXIV_vnth}
\end{figure*}   

Incorporated with the assumption of $T_i = T_e$, we used the measurements of $T_e$ made by \citet{Boe2020} during the 2017 TSE to calculate the nonthermal widths of Fe \textsc{x} and Fe \textsc{xiv} in Figure~\ref{fig:FeX_FeXIV_vnth}. The electron temperatures are inferred by the intensity ratio between Fe \textsc{xi} 789.2\,nm and Fe \textsc{xiv} 530.3\,nm, which is sensitive to temperatures ranging from 1--2\,MK. These two assumptions could introduce uncertainties because (1) $T_i$ may deviate from $T_e$ and (2) Fe \textsc{xi} and Fe \textsc{xiv} emissions may not originate from the same plasma structure along the LOS. The measured $T_e$ in the closed field regions ranges from 1.2\,MK to 1.4\,MK and does not show significant variations with height. Therefore, the distribution and variation of nonthermal widths in the off-limb corona look similar to that of total line widths. The nonthermal velocity $\xi$ appears to be minimal and constant in the AR, with a value of approximately 15--20\,km\,s$^{-1}$ Fe \textsc{x} $\xi$ exceeds 40\,km\,s$^{-1}$, suggesting that the assumption $T_i \approx T_e$ might not be valid in these open field regions.


\subsubsection{Wave or Turbulence-induced Nonthermal Motions}

Nonthermal broadening in coronal emission lines has been widely attributed to the propagation of Alfv\'enic waves, including torsional Alfv\'en and kink modes, as well as turbulence \citep[e.g.,][]{Seely1997}. Essentially, the product of Alfv\'en wave energy flux and the flux tube cross-section is proportional to $n_e^{1/2} \xi^2$ \citep[][]{Hassler1990}. Therefore, the wave propagation theory particularly favors the increase of line widths in the open fields, as $\xi \propto n_e^{-1/4}$ in the undamped regime \citep[e.g.,][]{Dolla2008,Banerjee2009}.   

Wave propagation may also result in narrower lines in closed fields. The wave energy flux leaked from the lower atmosphere should be relatively uniform across large scales, particularly in CHs and the QS, where the lower atmospheres are similar. Given $\xi \propto n_e^{-1/4}$, $\xi$ has to be greater in open fields with lower density. Additionally, the nonlinear Alfv\'en turbulence generated by the counter-propagating waves may dissipate the wave energy more efficiently in closed field regions \citep{vanderHolst2014}. 

In closed fields, relatively constant line widths may imply wave or turbulence dissipation in the lower corona. For instance, the constant nonthermal width shown in Figure~\ref{fig:FeX_FeXIV_vnth}, if solely attributed by Alfv\'en waves, would imply a decrease in the wave energy flux due to the density decrease with height. If the density drops by an order of magnitude from 1.0--1.5\,$R_\odot$ in the AR, as shown in Figure~\ref{fig:FeXIII_dens_FeXIV_pop}, the wave energy will decrease by 60--70\%. Considering typical values of $\xi \approx 15$\,km\,s$^{-1}$, $\log n_e \approx 9$, and $B \approx 20$\,G at the base of AR, the resulting Alfv\'en wave energy flux is approximately $10^6$\,erg\,cm$^{-2}$\,s$^{-1}$. This energy flux is insufficient to heat AR corona \citep[$10^7$\,erg\,cm$^{-2}$\,s$^{-1}$,][]{Withbroe1977}.  

On the other hand, even if the plasma is multithermal along the loop, the constant line width in the AR may still imply the existence of wave damping in ARs. To illustrate this, let's consider a simple case where $v_{\rm eff} \approx 30$\,km\,s$^{-1}$ in Fe \textsc{xiv} is contributed by $T_i = 2$\,MK and $\xi =17$\,km\,$s^{-1}$ at the base of the AR. If the waves are undamped and density drops by an order of magnitude, the new $\xi' = 10^{1/4}\xi \approx 30$\,km\,s$^{-1}$ would dominate the line broadening if $v_{\rm eff}$ remains nearly constant in ARs. Therefore, we should still consider the possibility of wave damping or other nonthermal motions not caused by wave propagation. 

\subsubsection{Other Nonthermal Motions}

Despite the fact that the possible wave damping in the closed fields \citep[e.g.,][]{Gupta2019}, an alternative explanation to relatively constant line widths is that the nonthermal width $\xi$ in closed fields does not primarily arise from Alfv\'en waves or turbulence. In fact, a component $\xi_\parallel$ parallel to the magnetic fields is often observed in on-disk ARs \citep[e.g.,][]{Brooks2016,Prabhakar2022}, which cannot be simply explained by the Alfv\'enic waves because the perturbation is perpendicular to the local magnetic field \citep[e.g.,][]{Shi2022}. For example, \citet{Asgari-Targhi2014} had to introduce a parallel component $\xi_\parallel$ with the Alfv\'en wave turbulence model to reproduce the $\xi$ observed by Hinode/EIS.

A few studies of the anisotropy of the nonthermal velocity confirm the existence of $\xi_\parallel$, but found different relationships between the two components $\xi_\parallel$ and $\xi_\perp$, including $\xi_\parallel > \xi_\perp$ \citep{Hahn2023} and $\xi_\parallel < \xi_\perp$ \citep{Hara1999}. The possible candidates to create $\xi_\parallel$ and the relatively constant nonthermal widths in closed fields include reconnection (e.g., jets and nanoflares) or slow mode waves \citep{Hahn2023}, but it is still difficult to reach a firm conclusion \citep{Brooks2016}. 

Additionally, \citet{Singh2002,Singh2006} proposed that the conduction between the warm and cold plasma within a single coronal loop could lead to a mixture of thermal and nonthermal motions below 1.3\,$R_\odot$. This conjecture is supported by observations showing a slight decrease in Fe \textsc{xiv} widths, an increase in Fe \textsc{x} width, and similar line widths between Fe \textsc{x} and Fe \textsc{xiv} from 1.2--1.3\,$R_\odot$ in the AR. Furthermore, other spectroscopic observations have suggested the existence of multi-strand (multi-thermal) loop cross-sections \citep[e.g.,][]{Aschwanden2013}. However, the scenario does not provide a clear explanation for the nature of the nonthermal motions and how they mix with each other within the multi-thermal plasma. A future study using coordinated spectroscopic observations in visible and EUV will provide new insights to explore this scenario. 



\subsection{Doppler Shifts in the Corona}

Significant Doppler shifts greater than 5\,km\,s$^{-1}$ were only observed in Fe \textsc{xiv} 530.3\,nm line in the northern and southern streamers, while the Fe \textsc{x} 637.4\,nm line does not show any notable Doppler shifts. Interestingly, a redshift of approximately 10\,km\,s$^{-1}$ observed in the northeast region only appeared in Fe \textsc{xiv}, suggesting the presence of bulk motions in the 2\,MK plasma. We argue that this redshift results from the plasma motions occurring at the boundary of open and closed-field structures, which is often suggested as the source region of the slow solar wind \citep[e.g.,][]{Antiochos2011}. Since the LOS component is approximately 10\,km\,s$^{-1}$, the total outflow velocity may easily exceed 10\,km\,s$^{-1}$. The upcoming 2024 TSE will be another unique opportunity to study the properties of these outflows, as the Solar Orbiter spacecraft \citep{Muller2020} will be in quadrature with the Earth during the eclipse. The stereoscopic observations of plasma flows in the lower corona will provide new insights into their source regions and evolution.

\section{Summary}\label{sec:sum}

We presented spectroscopic observations of the Fe \textsc{x} 637.4\,nm and Fe \textsc{xiv} 530.3\,nm visible forbidden lines during the 2017 TSE. Benefiting from the large FOV of 3PAMIS with its 4\,$R_\odot$ long slit,  we analyzed the line intensity, Doppler shifts, and broadening across various corona structures at the east limb up to 1.5\,$R_\odot$, including an AR, streamers, and a polar CH. We found distinct behaviors of the line widths between open and closed-field regions. In closed fields, the line widths are narrower, ranging from 20 to 30\,km\,s$^{-1}$, and relatively constant. In contrast, the line widths in open fields are broader ($>40$\,km\,s$^{-1}$) and increase with height between 1.0--1.3\,$R_\odot$. 

To complement our observations, we analyzed the observations from Hinode/EIS and CoMP, which provide consistent measurements of the line widths and support our findings. The EIS observations further allow us to measure the widths of other heavy ions in CH and QS regions. We discussed various underlying mechanisms, such as wave propagation, preferential heating, the LOS integration effect, and other nonthermal motions that may affect line widths in open- and closed-field structures. The differences in the width of spectral lines between various coronal structures suggest that wave heating is more dominant in open structures, while localized heating might occur in closed structures. 

This study highlights the unique advantages of TSE optical and near-infrared spectroscopy in line width measurements. First, line widths can be obtained at high altitudes with much shorter exposure times compared to conventional EUV spectroscopy. Second, the extended FOV enables simultaneous line width measurements in multiple open- and closed-field regions, providing a comprehensive overview of the global corona. Third, the minimal stray light levels and sky brightness during the totality, complemented by the smaller 3PAMIS instrumental widths, further reduce the uncertainty in measured line widths at great heliocentric distances.

Our study demonstrated the great potential of spectroscopic observations of visible forbidden lines during TSEs in advancing our knowledge of substantial problems, such as coronal heating and solar wind acceleration. For future observations, improvements in the effective area of the spectrograph will empower us to measure Doppler shifts and line widths in the higher corona, for instance, the Fe \textsc{x} width variation above 1.3\,$R_\odot$ in CHs. This region was seldom explored by UV observations of lines formed at $\sim$1\,MK (see Figure~\ref{fig:huge_compare}e) due to the sharp decrease in line intensity and instrumental stray light. Therefore, it would be of great interest to continue the spectroscopic observations of Fe \textsc{x} 637.4\,nm, and other visible forbidden lines during TSEs to shed light on the mysteries of the solar corona.


\section*{}
Funding for the DKIST Ambassadors program is provided by the National Solar Observatory, a facility of the National Science Foundation, operated under Cooperative Support Agreement number AST-1400405. EL acknowledges support from NASA grants 80NSSC20K0185 and 80NSSC22K0750. The 2017 total solar eclipse observations were supported by NASA grant NNX17AH69G and NSF grants AGS-1358239, AGS-1255894, and AST-1733542 to the Institute for Astronomy of the University of Hawaii. Financial support was provided to B.B. by the National Science Foundation under Award No. 2028173.

Hinode is a Japanese mission developed and launched by ISAS/JAXA, collaborating with NAOJ as a domestic partner, NASA and UKSA as international partners. Scientific operation of the Hinode mission is conducted by the Hinode science team organized at ISAS/JAXA. This team mainly consists of scientists from institutes in the partner countries. Support for the post-launch operation is provided by JAXA and NAOJ (Japan), UKSA (U.K.), NASA, ESA, and NSC (Norway). The CoMP data are courtesy of the Mauna Loa Solar Observatory, operated by the High Altitude Observatory, as part of the National Center for Atmospheric Research (NCAR). NCAR is supported by the National Science Foundation. The SDO data are courtesy of NASA/SDO and the AIA, EVE, and HMI science teams. CHIANTI is a collaborative project involving George Mason University, the University of Michigan (USA), University of Cambridge (UK) and NASA Goddard Space Flight Center (USA). YZ would like to thank G. de Toma and M. Galloy for providing the CoMP data with the latest calibration and helpful discussion on determining the zero point of the Doppler velocity. YZ also acknowledges B. van der Holst, T. Shi, A. Tei, X. Sun, Z. Yang, X. Liu, and H. Lu for all helpful discussions. The authors also acknowledge the anonymous referee for constructive comments and suggestions. 

%

\vspace{5mm}
\facilities{Hinode(EIS), SDO(AIA), CoMP\citep{CoMP1074Data,CoMP1079Data}}


\software{Numpy \citep{oliphant2006guide,van2011numpy}, Scipy \citep{2020SciPy-NMeth}, Astropy \citep{astropy:2013,astropy:2018,astropy:2022}, Specutils \citep{specutils2022}, Sunpy \citep{sunpy_community2020}, EISPAC, Matplotlib \citep{Hunter2007}, CHIANTI \citep{Dere1997,DelZanna2021}, SolarSoft \citep{Freeland2012}, OpenCV \citep{opencv_library}, num2tex\footnote{\url{https://github.com/AndrewChap/num2tex}}, cmcrameri \citep{Crameri2021}, Mathematica, ChatGPT, hissw \citep{hissw2023}. The \texttt{Jupyter} notebooks and \texttt{IDL} scripts for data reduction, analysis, and visualization are available on Zenodo \citep{CodeRepoZhuTSE} and GitHub.}



\appendix
\section{Data Calibration and Coalignment}\label{app:data_calib}
We performed data reduction and calibration of the raw CCD images through the following steps: (1) dark frame subtraction, (2) curvature correction, and (3) flat-fielding. Furthermore, we determined the instrument pointing, carried out the wavelength calibration, and measured the instrumental broadening of the spectrometer. 

We applied dark-frame subtraction to remove both the detector bias and dark current. We created master dark frames for each detector with exposure times of 1 s, 3 s, and 5 s, each obtained by averaging 10 dark frames with the same exposure time, after removing their hot pixels $>5\sigma$. These master dark frames were subsequently utilized to correct the CCD images with identical or similar exposure times. 

In addition, we corrected the curved spectral lines recorded by the detector. The correction of line curvature is crucial for various following calibrations, including flat-fielding, pixel binning along the $y$-axis, wavelength calibration, and fitting of line widths. To accomplish this, we employed emission from neutral hydrogen and helium calibration lamps in the laboratory to measure the curvature. 

\begin{figure}[htb]
    \centering
    \includegraphics[width=0.5\textwidth]{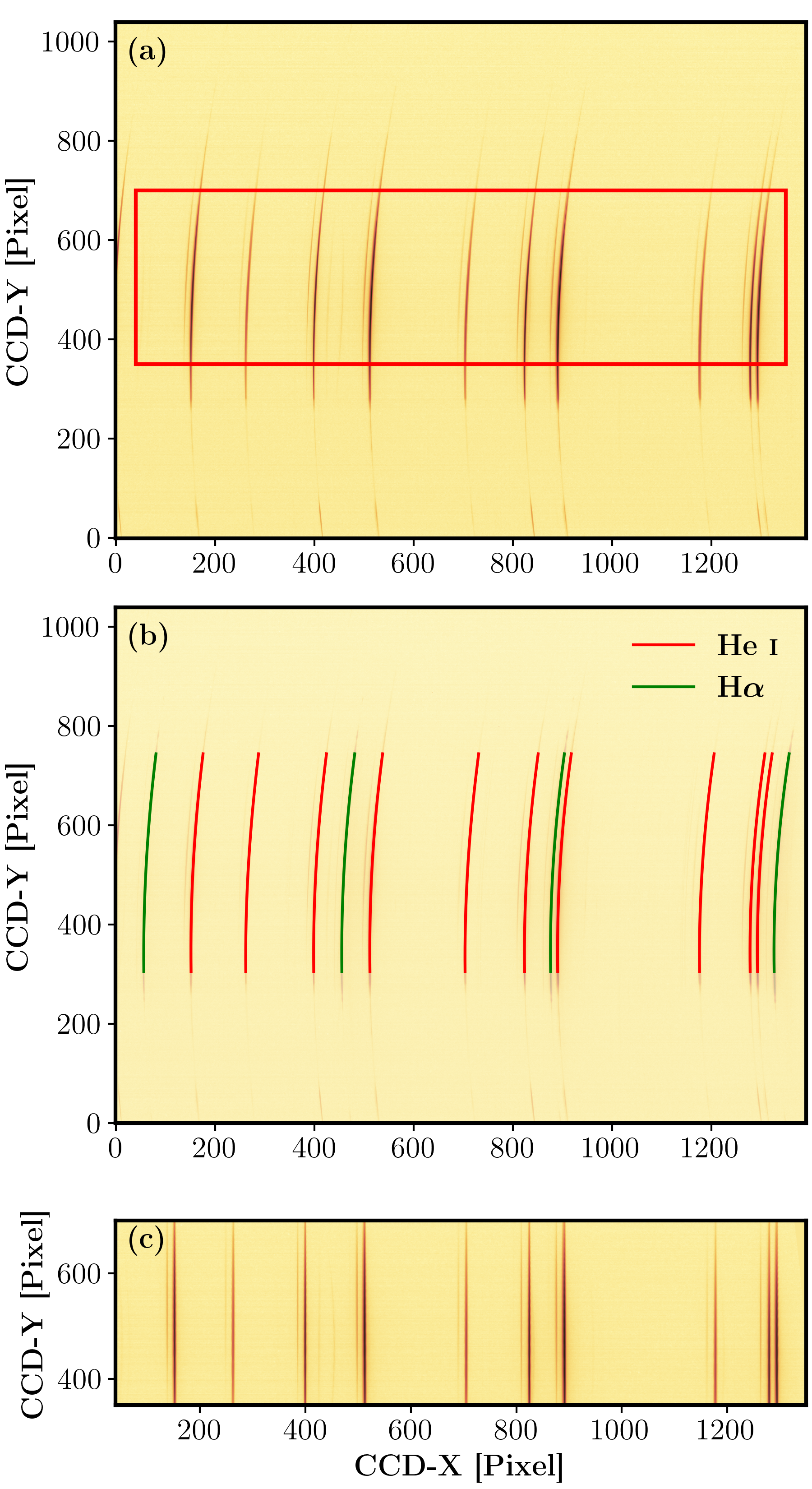}
    \caption{An example of curvature correction: (a) A CCD image showing curved neutral helium spectral lines on the detector. The red rectangle outlines the region shown in panel (c) where the curvature is corrected. (b) Fitting of curved neutral hydrogen and helium lines. (c) Curvature-corrected helium lines from the red rectangle in panel (a). Link to the \texttt{Jupyter} notebook creating this figure: \href{https://yjzhu-solar.github.io/Eclipse2017/ipynb_html/curvature_red_test.html}{\faGithub}.}
    \label{fig:appen_curve_corr}
\end{figure}

The measurement of the curvature was carried out in two steps: (1) We averaged every 5 pixels along the $y$-axis and fitted the line centroids at different CCD $y$-pixels using single-Gaussian fitting (see Figure~\ref{fig:appen_curve_corr}b). The shift along the $x$-axis was measured with respect to the line centroid at $y = 400$. (2) To extrapolate the shift from where calibration lines were located to the entire detector, we utilized a 2-D Chebyshev polynomial. The latter is a first-order polynomial in the $x$-direction and a second-order polynomial in the $y$-direction; it was used to fit the shift of all calibration lines at various parts of the detector. This approach was chosen to emulate the legacy Image Reduction and Analysis Facility (IRAF) and its user guide for slit spectroscopy \citep{Massey1992}. Then we interpolated each pixel along the $x$-axis to correct the curvature (Figure~\ref{fig:appen_curve_corr}c). All the images used in this study, except for the dark frames, were subjected to the curvature correction procedure.

\begin{figure}[htb]
    \centering
    \includegraphics[width=0.7\textwidth]{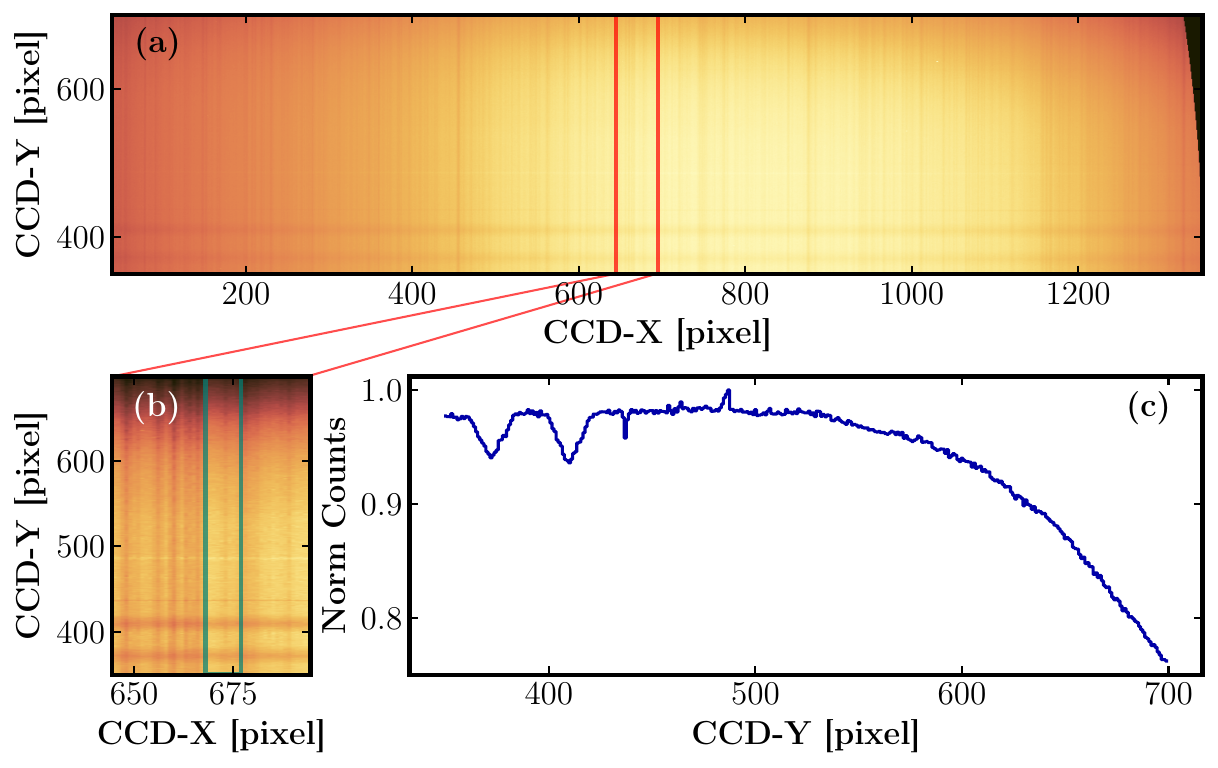}
    \caption{1-D flat field function of the red detector for 52nd-order Fe \textsc{x} 637.4\,nm line. (a) Curvature-corrected sky flat image. (b) Zoom-in sky flat image. (c) The 1-D flatfield function averaged between the two green vertical lines. Link to the \texttt{Jupyter} notebook creating this figure: \href{https://yjzhu-solar.github.io/Eclipse2017/ipynb_html/flatfield_red_curv_corr.html}{\faGithub}.}
    \label{fig:appen_flatfield}
\end{figure}

We performed flat-fielding corrections only along the $y$-direction for several reasons. First, the spectral lines are recorded along the $y$-axis of the detector. Second, the laboratory/dome flats were not used, as they were acquired when the slit was not evenly illuminated. Finally, the sky flats contain a considerable number of telluric lines (see Figure~\ref{fig:appen_flatfield}a). To obtain the 1-D flat-fielding function, we averaged the ``clean" sky flat images between telluric contamination. We carried out this procedure specifically at the regions of the detector where Fe \textsc{x} and Fe \textsc{xiv} lines were located. Figure~\ref{fig:appen_flatfield}c illustrates an example of the flat-fielding curve for the Fe \textsc{x} 637.4\,nm line at the 52nd order. It is important to note that the 1-D flat-fielding primarily corrects optical effects, such as vignetting, but does not correct the response differences of individual CCD pixels.

\begin{figure}[htb]
\gridline{\fig{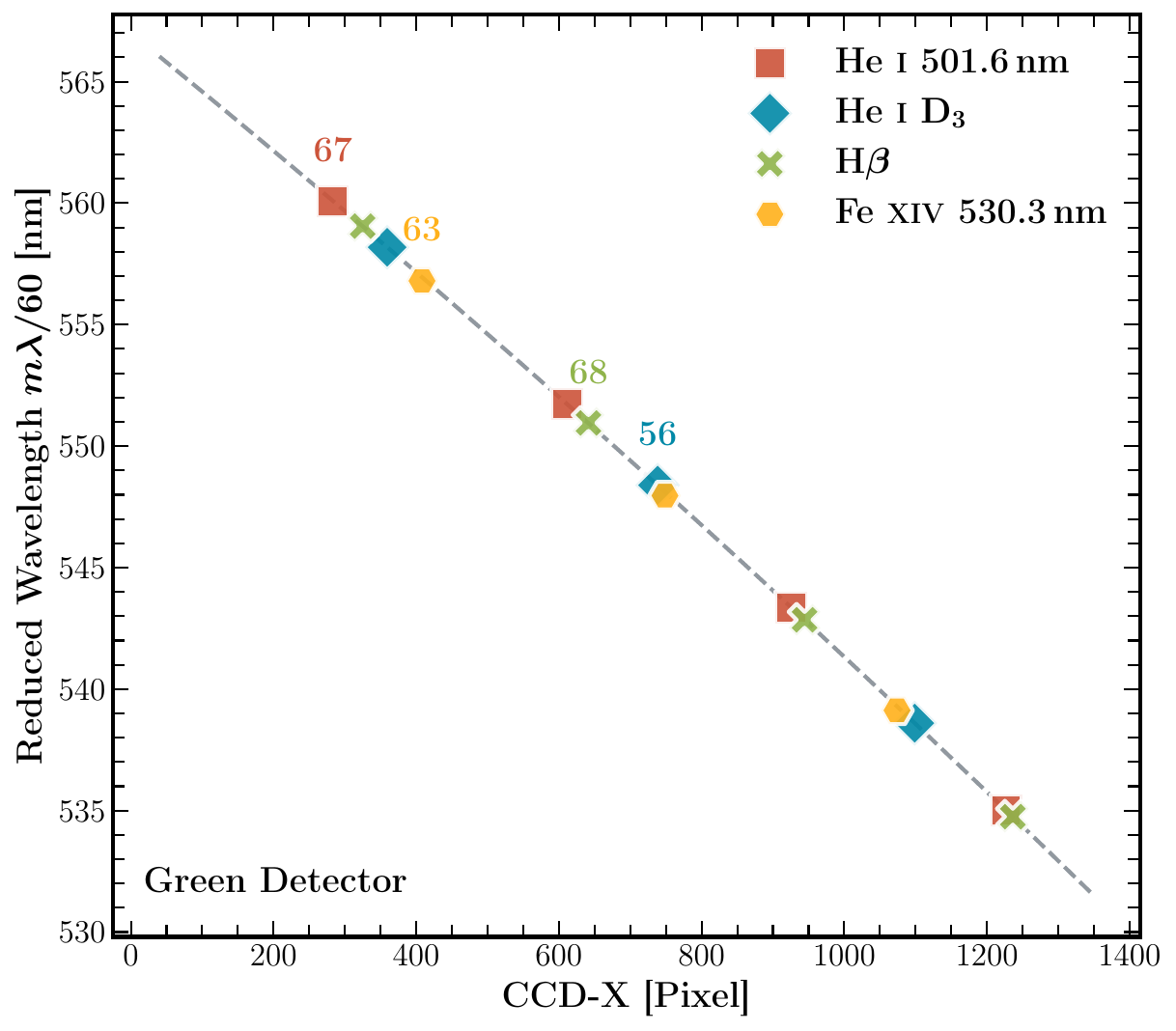}{0.49\textwidth}{(a)}
          \fig{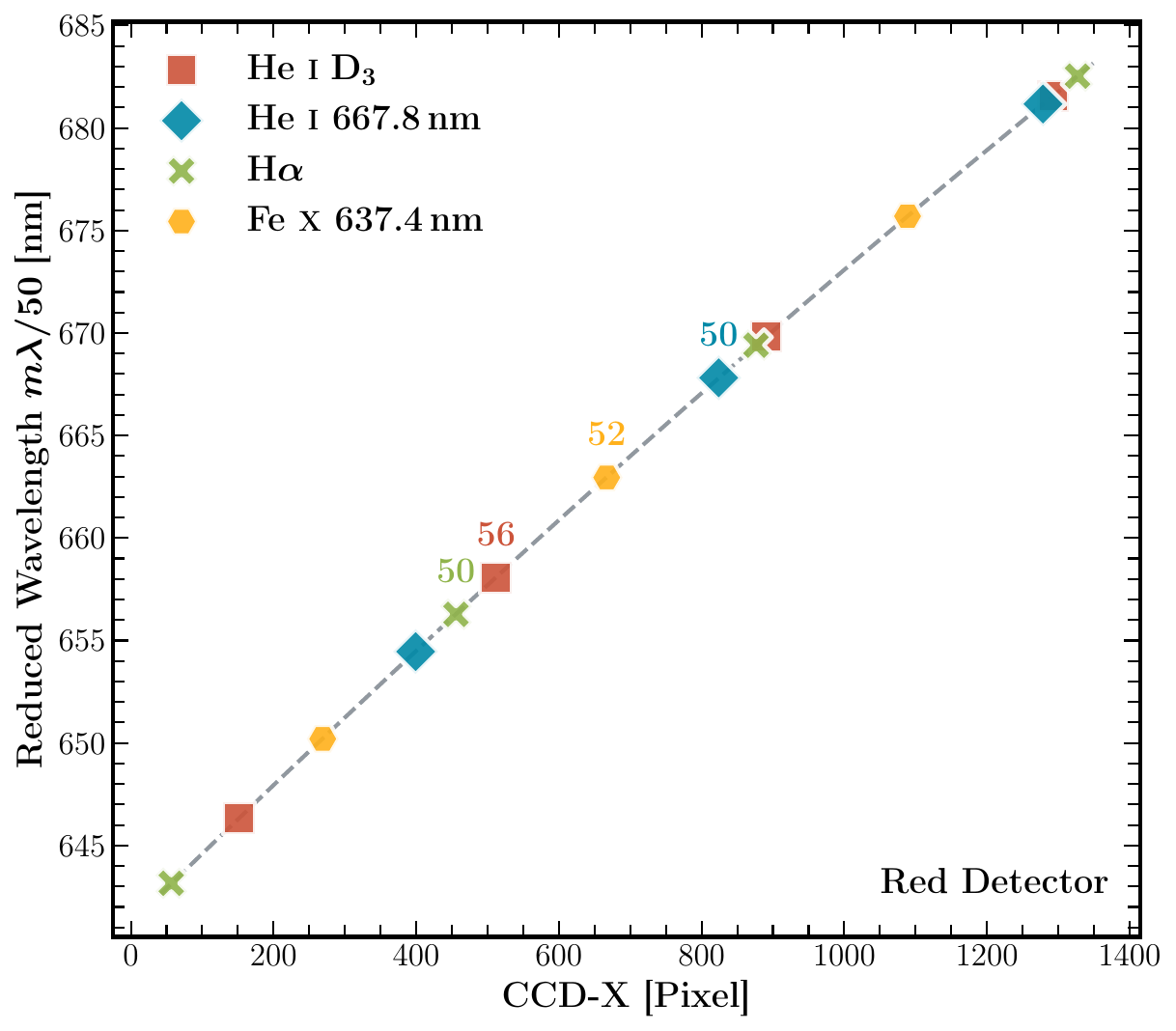}{0.49\textwidth}{(b)}}
    \caption{CCD $x$-pixel positions of various-order spectral lines and the relative wavelength calibration of the green (left) and red (right) detectors. The neutral hydrogen and helium lines are used in the wavelength calibration. The dashed curves show the quadratic fittings of wavelengths. The locations of the observed Fe \textsc{x} and Fe \textsc{xiv} lines during the eclipse are also shown. Links to \texttt{Jupyter} notebooks creating this figure: \href{https://yjzhu-solar.github.io/Eclipse2017/ipynb_html/wvl_calib_green_curv_corr.html}{\faGithub} (green) and \href{https://yjzhu-solar.github.io/Eclipse2017/ipynb_html/wvl_calib_red_curv_corr.html}{\faGithub} (red).}
    \label{fig:appen_relative_wvl_cal}
\end{figure}

We performed the wavelength calibration in two steps: (1) a relative wavelength calibration using laboratory hydrogen and helium calibration lines, followed by (2) an absolute wavelength calibration using chromospheric hydrogen and helium emission at the limb. Figure~\ref{fig:appen_relative_wvl_cal} shows the relative wavelength calibration of the green and red detectors. Specifically, H$\beta$, He \textsc{i} 501.6\,nm, and He \textsc{i} $\mathrm{D_3}$ are used for the green detector and H$\alpha$, He \textsc{i} $\mathrm{D_3}$, and He \textsc{i} 667.8\,nm are used for the red detector. The He \textsc{i} $\mathrm{D_3}$ line at 587.6\,nm can be observed in both detectors because it is close to the wavelength limit of the dichroic mirror. To derive the wavelength scale, we adopted a second-order polynomial to fit the NIST air wavelengths of neutral hydrogen and helium lines. The wavelength scales at the detector center are approximately 0.025\,$\mathrm{nm\,px^{-1}}$ (62nd order, green) and 0.030\,$\mathrm{nm\,px^{-1}}$ (52nd order, red). 

A significant deviation from the laboratory wavelength scale was found in the totality images, likely resulting from slight perturbations of the optics during transportation and deployment. Therefore, we carried out an additional absolute wavelength calibration using chromospheric lines at the limb to correct the reference wavelength. 

\begin{figure}[htb]
    \centering
    \includegraphics[width=0.5\textwidth]{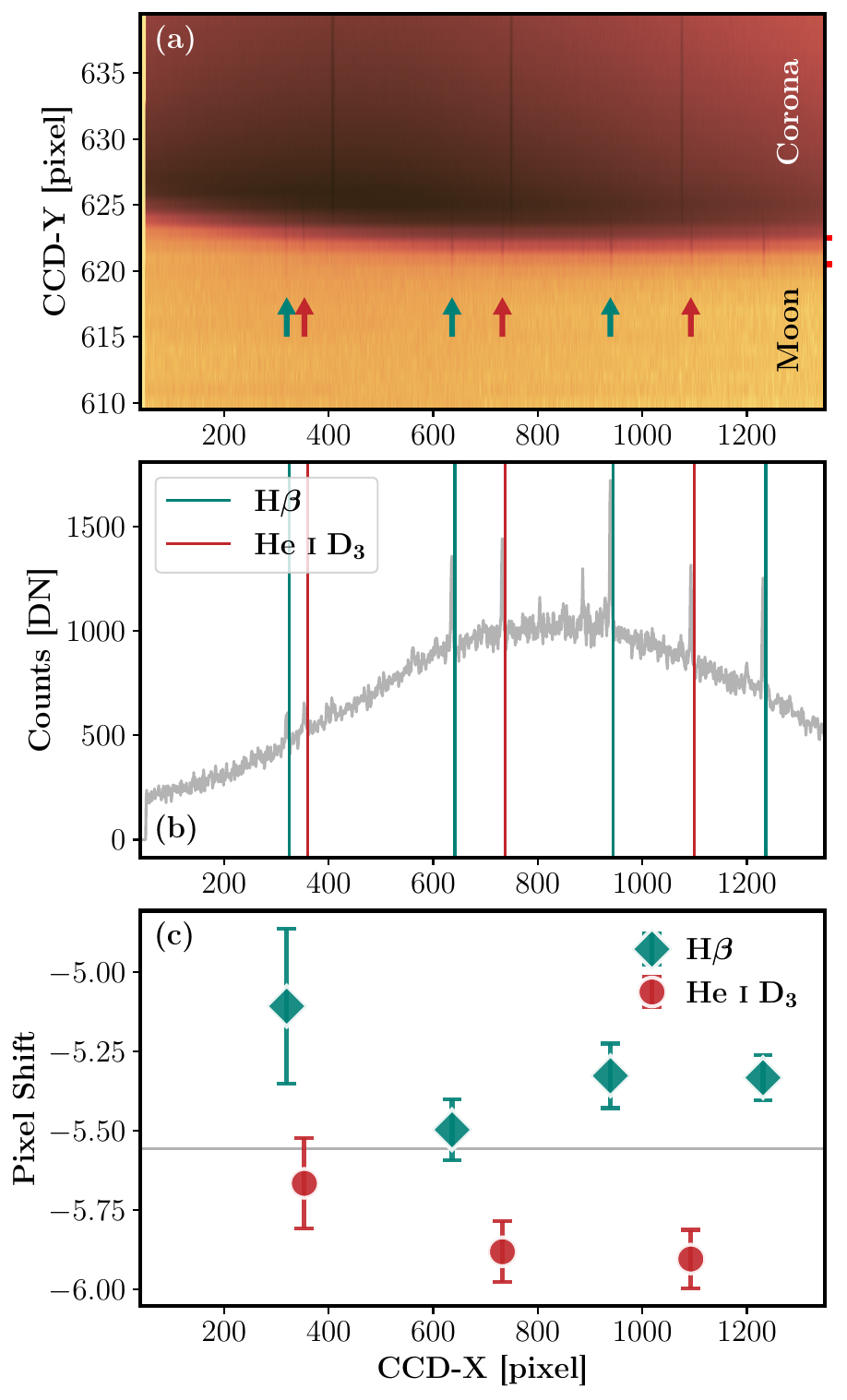}
    \caption{Absolute wavelength calibration of the green detector. (a) Chromospheric H$\beta$(green arrow) and He \textsc{i} (red arrow) emissions between the lunar disk and corona. (b) The average chromospheric spectrum (grey curve) between the two red ticks in panel (a). The vertical green and red lines indicate the line centroids of laboratory H$\beta$ and He \textsc{i}. (c) The shift between the chromospheric lines and the laboratory lines at different orders. The horizontal grey line indicates the average pixel shift for the absolute wavelength calibration. Link to the \texttt{Jupyter} notebook creating this figure: \href{https://yjzhu-solar.github.io/Eclipse2017/ipynb_html/abs_wvl_calib_chromo.html}{\faGithub}.}
    \label{fig:appen_abs_wvl_cal}
\end{figure}

Figure~\ref{fig:appen_abs_wvl_cal} displays the absolute wavelength calibration of the green detector. We found a shift of approximately 5 pixels along the $x$-direction between the chromospheric lines and the same lines measured in the laboratory. We utilized the average pixel shift at different orders to update the reference wavelength. The average shifts are found to be -5.56 pixels for the green detector and -1.17 pixels for the red detector. Moreover, the spread of the pixel shift in Figure~\ref{fig:appen_abs_wvl_cal}c allowed an estimation of the uncertainty in wavelength calibration, which was about 1/3 pixel ($\approx 0.008$\,nm or $4.5\,\mathrm{km\,s^{-1}}$ at 530\,nm). Additionally, we chose the median value of the Doppler shift measured in the FOV as the zero point velocity to remove the solar rotation.

In addition to the wavelength calibration, we adopted the neutral hydrogen and helium lines to measure the instrumental broadening. As the thermal and nonthermal broadening of these calibration lines is negligible, the widths of calibration lines provided a direct measurement of the instrument widths. We noticed that the interpolation to correct the line curvature might affect the fitted line widths because the calibration lines are very narrow and usually only sampled by 4-5 pixels in the $x$-direction. 

\begin{figure}[htb]
    \centering
    \includegraphics[width=0.7\textwidth]{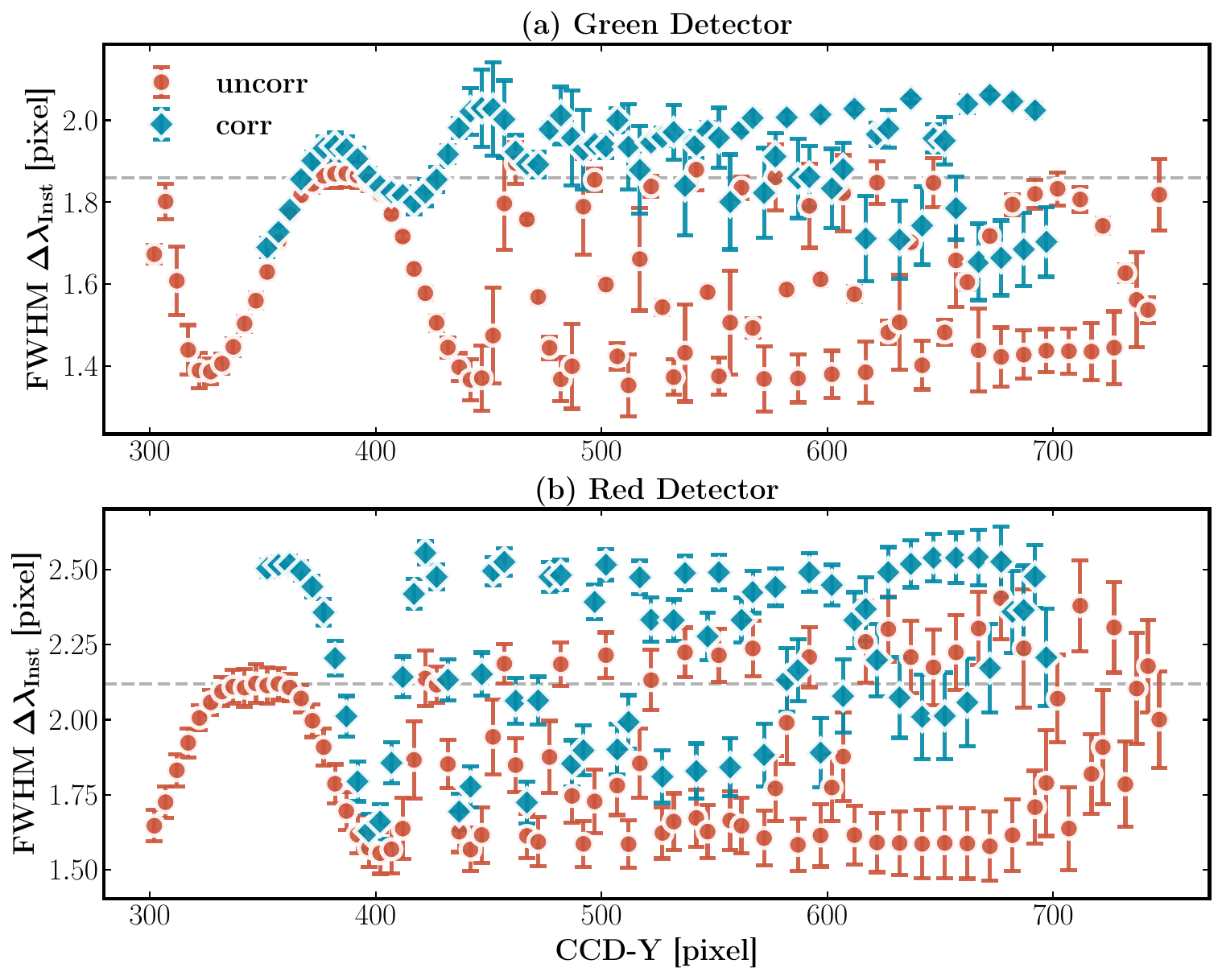}
    \caption{Line widths of the narrow neutral hydrogen and helium lines as a function of CCD $y$-pixel position in the green (a) and red (b) detectors. The blue diamonds represent the curvature-corrected widths, and the red dots are for the uncorrected widths. The dashed horizontal lines indicate the instrumental widths used in this study. Link to the \texttt{Jupyter} notebook creating this figure: \href{https://yjzhu-solar.github.io/Eclipse2017/ipynb_html/plot_instwidth.html}{\faGithub}.}
    \label{fig:appen_instwidth}
\end{figure}

To address this concern, we compared the curvature-corrected and uncorrected widths as a function of CCD $y$-pixel in Figure~\ref{fig:appen_instwidth}. We found that the two line widths agree with each other when the curvature is negligible ($y\approx 400$). However, at locations where lines are curved, the uncorrected widths fluctuate from 1.4 to 1.9 pixels (green) and 1.5 to 2.1 pixels (red), which might be attributed to the insufficient sampling of the line profile. Notably, the curvature correction failed to remove the fluctuation, particularly for the red detector, where the curvature-corrected widths vary from 1.6 pixels to 2.5 pixels.

Finally, we selected the region where the uncorrected widths do not vary significantly to measure the instrumental widths, at $y\sim 380$ for the green detector and $y\sim 350$ for the red detector. The instrumental widths of $\Delta \lambda_{\rm inst, green} = 1.86$\,px and $\Delta \lambda_{\rm inst, red} = 2.12$\,px are used in this study, with an uncertainty of approximately $20-30\%$. We estimated the uncertainties of instrumental widths as $\sigma_{\rm inst} = 0.4$\,px (green) and $\sigma_{\rm inst} = 0.5$\,px (red), based on the spread of values depicted in Figure~\ref{fig:appen_instwidth}. The instrumental widths are removed by 
\begin{equation}
    \Delta \lambda_{\rm true} = \sqrt{\Delta \lambda_{\rm fit}^2 - \Delta \lambda_{\rm inst}^2}
\end{equation}
where $\Delta \lambda_{\rm true}$ is the deduced true line width. We also propagated the uncertainty by 
\begin{equation}
    \sigma_{\rm true} = \left(\frac{\lambda_{\rm fit}^2}{\lambda_{\rm true}^2}\sigma_{\rm fit}^2 + \frac{\lambda_{\rm inst}^2}{\lambda_{\rm true}^2}\sigma_{\rm inst}^2\right)^{1/2}
\end{equation}
where $\sigma_{\rm true}$ is the uncertainty of $\Delta \lambda_{\rm true}$, $\sigma_{\rm fit}$ denotes the fitting uncertainty. 

Regrettably, the 3PAMIS FITS headers do not include any pointing information since the instrument was set up on-site manually, only a few days prior to the eclipse. Therefore, we had to rely on the white light images taken by the context camera to determine the pointing. Although the context camera did not record the time of observation, this information was available in 3PAMIS FITS headers. Hence, we manually compared the 3PAMIS images taken at the onset of totality with the context images to determine the reference time. 

\begin{figure}[htb]
    \centering
    \includegraphics[width=0.7\textwidth]{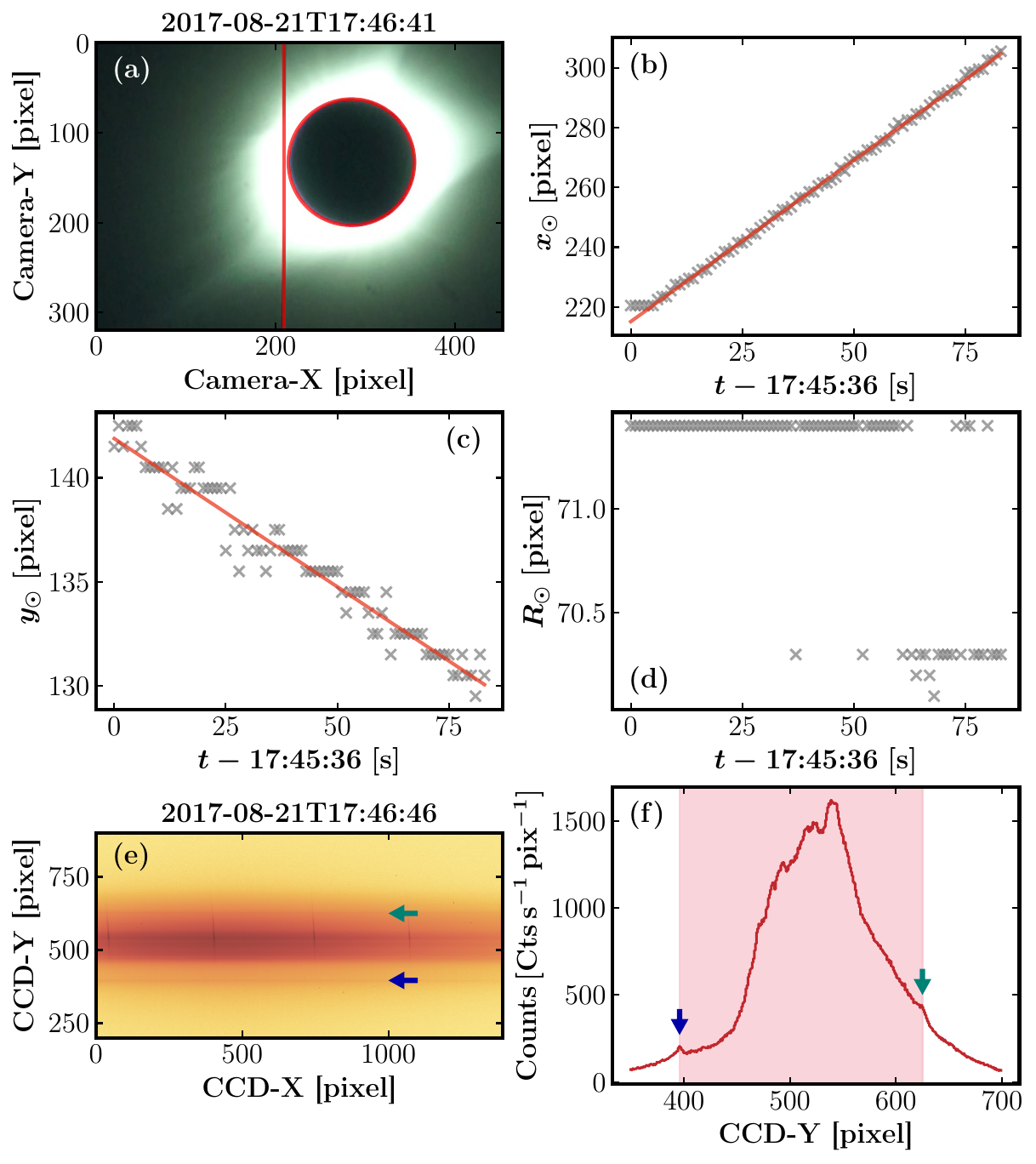}
    \caption{Determining the spectrograph pointing: (a) Fitting the lunar limb on context images. The red cycle highlights the lunar limb, and the red vertical line represents the position of the slit. (b) and (c) Linear fitting of the motion of the disk center on context images. (d) Variation of the fitted disk radius. (e) An off-limb CCD image. The blue and green arrows indicate limb emission leaking from the slit mirror. (f) CCD counts averaged along the $x$-axis, with the arrows being the same as Panel (e). The shaded pink area between the two arrows is used to calculate the spatial sampling of the detector. Link to the \texttt{Jupyter} notebook creating this figure: \href{https://yjzhu-solar.github.io/Eclipse2017/ipynb_html/pointing_coalignment.html}{\faGithub}.}
    \label{fig:appen_pointing}
\end{figure}

To begin, we determined the slit position relative to the solar disk. As the pointing of the slit was fixed, our task involved measuring the motion of the Sun in the context images. To accomplish this, we adopted the circle Hough Transform method \citep{Duda1972} available in OpenCV to detect the lunar limb as a circular feature in the images (see Figure~\ref{fig:appen_pointing}a). Subsequently, we performed linear fitting on the $x$ and $y$-coordinates of the lunar disk when the Sun crossed the slit (Figure~\ref{fig:appen_pointing}b and c) to measure the velocities of the Sun $v_{\odot,x}$ and $v_{\odot,y}$. To convert these velocities in pixels into arcsecs, we compared the radius of the lunar disk (approximately $71.4$ pixels, see Figure~\ref{fig:appen_pointing}d), with the lunar disk size 976\arcsec\ reported in \citet{Boe2020}.

We took advantage of the semi-transparent slit mirror to measure the spatial scale $\Delta y$ of the detectors. Figure~\ref{fig:appen_pointing}e displays a CCD image captured when the slit was pointed to the off-limb. Two faint horizontal lines can be identified on the image, which are caused by the dispersion of the limb image leaking from the semi-transparent slit mirror. We used the positions of these lines to derive the spatial scale $\Delta y$ of two detectors and the position of the disk center $y_c$ on the detector. The spatial scales $\Delta y$ of the two detectors are measured to be 8\farcs26\,$\mathrm{px^{-1}}$ (green) and 8\farcs37\,$\mathrm{px^{-1}}$ (red).

The final crucial parameter is the angle $\alpha$ between the slit and the solar north--south direction. The slit was slightly tilted from solar northwest to southeast as depicted in Figure~\ref{fig:appen_pointing}a. Initially, we compared the locations of streamers in the white light context images with the reference eclipse images from \citet{Boe2020}. However, because the inner corona was saturated in the context images, this approach only yielded an angle of approximately 30\arcdeg. Therefore, we manually compared the Fe \textsc{xiv} intensity observed by 3PAMIS with Fe \textsc{xiv} narrowband images to obtain a better estimation of the angle $\alpha = 27\fdg5$. 

Finally, the transformation from detector pixel at $y$-pixel $y_{i}$ taken at time $t_i$ to the helioproject coordinates $(\theta_x, \theta_y)$ in arcsec is given by 

\begin{equation}
    \begin{bmatrix}
    \theta_x \\ \theta_y
    \end{bmatrix}
    = 
    \begin{bmatrix}
        \cos\alpha & \sin\alpha \\
        -\sin\alpha & \cos\alpha 
    \end{bmatrix}
    \begin{bmatrix}
        (t_i - t_0)v_{\odot,x} + \theta_{x,0} \\
        (y_i - y_{c})\Delta y + (t_i - t_0)v_{\odot,y}
    \end{bmatrix}
\end{equation}
where $\alpha = 27\fdg5$ is the slit tilting angle. $v_{\odot,x}$ and $v_{\odot,y}$ are the velocities of the Sun in arcsec captured by the context camera. $t_0 = $UT17:46:38 corresponds to the reference time when the slit first pointed to off-limb. $\theta_{x,0}$ denotes the distance between the slit and disk center at $t_0$ in arcsec. $y_c$ represents the disk center position on the detector at $t_0$. 

It is important to acknowledge that the method used to determine the instrumental pointing has certain limitations. First, the circle Hough Transform method used to detect the lunar limb has a precision of 1/2 pixel, which translates to about 6\farcs7. Second, the time of observation recorded in the 3PAMIS FITS header has a limited precision of 1 sec. Given that the Sun moved nearly perpendicular to the slit at a speed of about 15\arcsec\,$\mathrm{s^{-1}}$, an inaccurate time of observation may result in an uncertainty of $\sim 10$\arcsec\  perpendicular to the slit. Third, the slit tilt angle $\alpha$ determined by comparing Fe \textsc{xiv} intensity with narrowband images cannot achieve better results than the spatial scale of the detector, which is $\sim 8\arcsec$. This limitation could be more significant towards the two ends of the slit due to the focus on comparing the features in the AR. Additionally, the rotation of the slit mixes the uncertainty along and perpendicular to the slit. Overall, we estimated that the pointing used in this study might have an uncertainty up to 20--30\arcsec. 

No radiometric calibration was performed due to the lack of laboratory light sources and the multi-order design of the spectrometer. However, we still estimated the uncertainty in each pixel, assuming the photon shot noise follows the Poisson statistics:
\begin{equation}
    \sigma_D   = \sqrt{D_0 + \sigma_{0}^2}
\end{equation}
where $D_0$ is the total count in data number (DN) after the dark frame subtraction, and $\sigma_0$ is the combination of CCD readout noise and dark current noise estimated from the standard deviation of the master dark frames. We note that $\sigma_D$ only represents the relative magnitude of the uncertainty in each pixel since the absolute magnitude of the photon shot noise is the square root of the photon counts or photon electron counts. The non-linear least square routine to fit the Gaussian profiles will automatically rescale these uncertainties to reach a unity $\chi^2$. 

\section{Photon Redistributions of Constant Continuum}\label{app:scattering}
Physically, it is trivial, to some extent, that the flat continuum resonantly scattered by a Gaussian absorption profile leads to another Gaussian profile. To provide a detailed proof in this Section, we consider the local emissivity $ \epsilon(\nu, \boldsymbol{\hat{n}})$ at a frequency $\nu$ in the direction of $\boldsymbol{\hat{n}}$ caused by the incident emission is proportional to 
\begin{equation}
    \epsilon(\nu, \boldsymbol{\hat{n}}) \propto \frac{1}{4\pi} \int \mathrm{d}\Omega' \int_0^\infty \mathrm{d}\nu' R(\nu', \boldsymbol{\hat{n}}';\nu, \boldsymbol{\hat{n}}) I(\nu',\boldsymbol{\hat{n}}') 
\end{equation}
where the integral over the differential solid angle $\mathrm{d}\Omega'$ describes the scattering of incoming photons from different directions. $R(\nu', \boldsymbol{\hat{n}}';\nu, \boldsymbol{\hat{n}})$ is the photon redistribution function in the observer's frame, which describes the probability to scatter an incoming photon at frequency $\nu'$ and along direction $ \boldsymbol{\hat{n}}'$ to a new frequency $\nu$ and direction $\boldsymbol{\hat{n}}$, and $I(\nu',\boldsymbol{\hat{n}}')$ is the incoming photospheric radiation intensity. Consider the scattering from a constant (flat) continuum $I(\nu',\boldsymbol{\hat{n}}') = I(\boldsymbol{\hat{n}}')$, we have 
\begin{equation}
    \epsilon(\nu, \boldsymbol{\hat{n}}) \propto \frac{1}{4\pi} \int I(\boldsymbol{\hat{n}}') \mathrm{d}\Omega' \int_0^\infty \mathrm{d}\nu' R(\nu', \boldsymbol{\hat{n}}';\nu, \boldsymbol{\hat{n}})  
\end{equation}
Let's first deal with the integral over the incident frequency 
\begin{equation}
    \mathcal{I}_1 = \int_0^\infty R(\nu', \boldsymbol{\hat{n}}';\nu, \boldsymbol{\hat{n}})  \mathrm{d}\nu' \label{eq:B6}
\end{equation}
Assuming the scattering happens between two sharp energy levels, neglecting the natural broadening \citep[Case I,][]{Mihalas1978}, the analytical form of the photon redistribution function can be written as \citep{Cranmer1998,Gilly2020}
\begin{equation}
     R(\nu', \boldsymbol{\hat{n}}';\nu, \boldsymbol{\hat{n}}) = \frac{g(\Theta)}{\pi \beta (\Delta \nu)^2} \exp\left(-\zeta'^2 \right) \exp\left[-\left(\frac{\zeta - \alpha \zeta'}{\beta} \right)^2 \right]
\end{equation}
where $\Theta = \langle \boldsymbol{\hat{n}}',\boldsymbol{\hat{n}}\rangle$, $\alpha \equiv \cos \Theta$, and $\beta \equiv \sin \Theta$. $g(\Theta)$ is the angular distribution phase function. $\Delta \nu = \nu_0 v_{\rm eff}/c$ represents the local effective velocity in frequency units, where $\nu_0$ is the rest frequency of the spectral line. Additionally, $\zeta$ and $\zeta'$ are dimensionless frequency displacements defined by
\begin{align}
    \zeta &\equiv \frac{\nu - \nu_0}{\Delta \nu} - \frac{\boldsymbol{u}\cdot \boldsymbol{\hat{n}} }{v_{\rm eff}} = \frac{\nu - \nu_0\left(1 + u_{\rm LOS}/c\right)}{\Delta \nu} \\
    \zeta' &\equiv \frac{\nu' - \nu_0}{\Delta \nu} - \frac{\boldsymbol{u}\cdot \boldsymbol{\hat{n}}' }{v_{\rm eff}}  
\end{align}
where $\boldsymbol{u}$ denotes the local bulk velocity and $\boldsymbol{u}\cdot \boldsymbol{\hat{n}}$ is the local LOS velocity $u_{\rm LOS}$. Replacing $\mathrm{d}\nu'$ with $\Delta \nu \mathrm{d} \zeta'$, the integral in Equation~(\ref{eq:B6}) can be written as 
\begin{align}
  \mathcal{I}_1 &= \frac{g(\Theta)}{\pi \beta \Delta \nu} \int_{-\nu_0/\Delta \nu - \boldsymbol{u}\cdot \boldsymbol{\hat{n}}'/v_{\rm eff}}^\infty \exp\left(-\zeta'^2 \right) \exp\left[-\left(\frac{\zeta - \alpha \zeta'}{\beta} \right)^2 \right] \mathrm{d}\zeta' \nonumber \\
  &\approx \frac{g(\Theta)}{\pi \beta \Delta \nu} \int_{-\infty}^\infty \exp\left(-\zeta'^2 \right) \exp\left[-\left(\frac{\zeta - \alpha \zeta'}{\beta} \right)^2 \right] \mathrm{d}\zeta' \nonumber \\
  &= \frac{g(\Theta)}{\sqrt{\pi} \Delta \nu} \exp\left(-\zeta^2 \right)  
\end{align}
note that we also used $v_0 \gg \Delta \nu$ and $\alpha^2 + \beta^2 = 1$. Notably, $\mathcal{I}_1$ is a Gaussian profile, which does not depend on $\boldsymbol{\hat{n}}'$ anymore. Thus, the local emissivity is 
\begin{equation}
    \epsilon(\nu, \boldsymbol{\hat{n}}) \propto \frac{\mathcal{I}_1}{4\pi} \int I(\boldsymbol{\hat{n}}') \mathrm{d}\Omega'
\end{equation}
where the integral over the solid angle is a scale factor due to the limb darkening. Therefore, the local emissivity is still a Doppler-shifted Gaussian function broadened by the local effective velocity $v_{\rm eff}$. 

\section{Comparison between Orders} \label{appen:orders}

\begin{figure*}[htb]
    \centering
    \includegraphics[width=\linewidth]{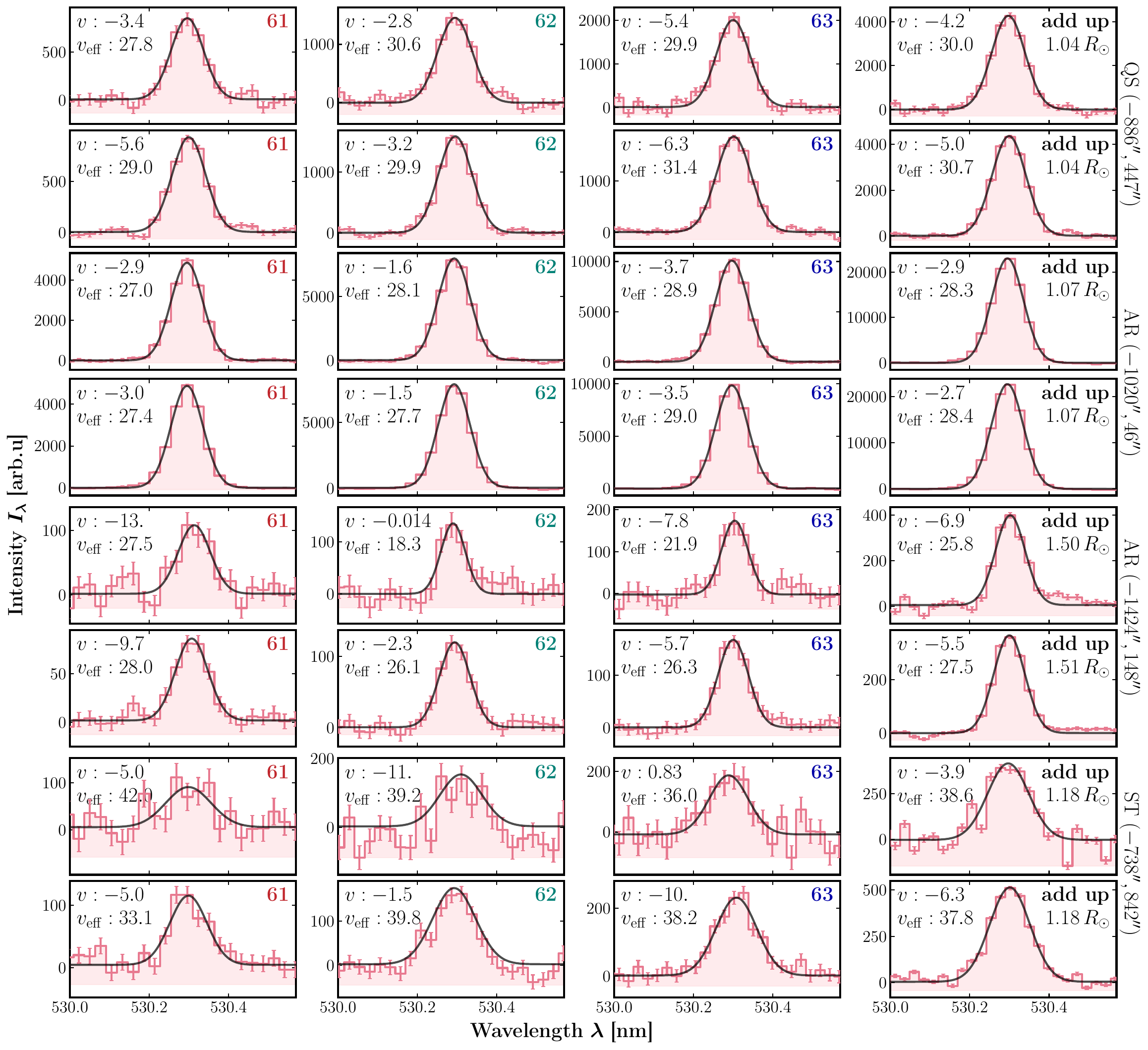}
    \caption{Comparison between Fe \textsc{xiv} 530.3\,nm profiles at the 61st, 62nd, and 63rd orders in various coronal structures and heights: the QS, AR, and streamer (ST). The profiles in each structure are depicted in two rows of subplots: the first row shows the original profiles, while the second row displays the 5-pixel averaged profiles. The last row shows the sum of three different orders. The fit Doppler velocity $v$ and effective velocity $v_{\rm eff}$ are measured in the units of km\,s$^{-1}$. Link to the \texttt{Jupyter} notebook creating this figure: \href{https://yjzhu-solar.github.io/Eclipse2017/ipynb_html/off_limb_intensity_map_ext_view_combine_3orders.html}{\faGithub}.}
    \label{fig:FeXIV_3orders}
\end{figure*}

One of the advantages of 3PAMIS is its capability to work at multiple orders, allowing for measurements of line profiles multiple times. However, in this study, we have only adopted the most prominent order of Fe \textsc{x} and Fe \textsc{xiv} lines. To improve the S/N, a natural approach is to combine line profiles in different orders to make the best use of collected photons.

In Figure~\ref{fig:FeXIV_3orders}, we present examples of Fe \textsc{xiv} line profiles at various orders obtained from different off-limb locations. In addition, we interpolated the line profiles at various orders to the same wavelength scale and summed these profiles together. This allows us to assess the performance of the instrument across different orders and examine the combined profiles.

In general, the fit results of Fe \textsc{xiv} line profiles at different orders are similar, especially when the S/N is high or after applying the 5-pixel average. In the brightest region of the AR, the differences in the Doppler shift and $v_{\rm eff}$ among various orders are typically less than 2\,km\,s$^{-1}$, which might be caused by the uncertainty of the absolute wavelength calibration. However, in the regions where the S/N is low, such as the AR at 1.5\,$R_\odot$, and the fainter streamer, the difference in the Doppler velocity measurements in different orders may exceed 5--10\,km\,s$^{-1}$. 

The averaging along the slit and the summation of different orders greatly improve the S/N and result in more Gaussian-like profiles. The combination of various orders generally averages the Doppler shifts and line widths of individual profiles. We only used the strongest orders in the data analysis because the detectors were primarily focused on the strongest orders, which could potentially make the profiles asymmetric at weaker orders. However, the comparison highlights the potential of taking full advantage of multiple orders in future observations with caution, especially when the flatfield is properly made. 


\bibliography{ms}{}
\bibliographystyle{aasjournal}



\end{document}